\documentclass{aa}

\usepackage[varg]{txfonts}
\usepackage{graphicx}
\usepackage{color} 
\usepackage{amsmath}
\usepackage{multirow} 
\bibpunct{(}{)}{;}{a}{}{,} 

\begin{document}

\title{Temperature profiles of young disk-like structures:}
\subtitle{The case of IRAS 16293A}

\author{Merel L.R. van 't Hoff\inst{1}
\and Ewine F. van Dishoeck\inst{1,2}
\and Jes K. J{\o}rgensen\inst{3}
\and Hannah Calcutt\inst{4}}

\institute{Leiden Observatory, Leiden University, P.O. box 9513, 2300 RA Leiden, The Netherlands\\e-mail: \texttt{vthoff@strw.leidenuniv.nl}
\and Max-Planck-Institut f\"ur Extraterrestrische Physik, Giessenbachstrasse 1, 85748 Garching, Germany 
\and Niels Bohr Institute, University of Copenhagen, {\O}ster Voldgade 5--7, 1350 Copenhagen K., Denmark
\and Department of Space, Earth and Environment, Chalmers University of Technology, 41296, Gothenburg, Sweden }

\date{}

\abstract {Temperature is a crucial parameter in circumstellar disk evolution and planet formation, because it governs the resistance of the gas to gravitational instability and sets the chemical composition of the planet-forming material.} 
{We set out to determine the gas temperature of the young disk-like structure around the Class 0 protostar IRAS 16293--2422A.} 
{We use Atacama Large Millimeter/submillimeter Array (ALMA) observations of multiple H$_2$CS $J=7-6$ and $J=10-9$ lines from the Protostellar Interferometric Line Survey (PILS) to create a temperature map for the inner $\sim$200 AU of the disk-like structure. This molecule is a particularly useful temperature probe because transitions between energy levels with different $K_{\rm{a}}$ quantum numbers operate only through collisions.}
{Based on the H$_2$CS line ratios, the temperature is between $\sim$100--175 K in the inner $\sim$150 AU, and drops to $\sim$75 K at $\sim$200 AU. At the current resolution (0.5$^{\prime\prime} \sim$70 AU), no jump is seen in the temperature at the disk-envelope interface.}
{The temperature structure derived from H$_2$CS is consistent with envelope temperature profiles that constrain the temperature from 1000 AU scales down to $\sim$100 AU, but does not follow the temperature rise seen in these profiles at smaller radii. Higher angular resolution observations of optically thin temperature tracers are needed to establish whether cooling by gas-phase water, the presence of a putative disk or the dust optical depth influences the gas temperature at $\lesssim$100 AU scales. The temperature at 100 AU is higher in IRAS 16293A than in the embedded Class 0/I disk L1527, consistent with the higher luminosity of the former.}


\maketitle


\section{Introduction}

Disks form around young stars due to conservation of angular momentum during the gravitational collapse of a dense core \citep{Cassen1981} and many disks have been reported around T Tauri and Herbig stars (Class II objects, e.g., \citealt{Andrews2005,Ansdell2016}). Disk-like structures are also observed in continuum emission toward embedded protostars \citep[Class 0 and II, e.g.,][]{Jorgensen2009,Tobin2016}, but Keplerian rotation has been established for only a handful of the youngest Class 0 sources \citep{Tobin2012,Murillo2013,Lindberg2014,Codella2014,Yen2017}. The formation of rotationally supported disks thus remain poorly constrained \citep[see e.g.,][for a review]{Li2014}, as well as their physical and chemical structure. However, it is now becoming clear that planet formation already starts in these young disks. The mass of mature protoplanetary disks seems too low to form the planetary systems that we observe \citep[e.g.,][]{Ansdell2016,Manara2018}, while younger disks are more massive \citep{Tychoniec2018,Williams2019}. This suggests that by the Class II stage, material has grown to larger bodies that can not be observed at (sub-)mm wavelengths. The first steps of grain growth have indeed been observed in disks that have not yet fully emerged from their envelope \citep{Kwon2009,Jorgensen2009,Pagani2010,Foster2013,Miotello2014,Harsono2018}. Young embedded disks thus provide the initial conditions for planet formation.   

An important unknown for young disks is the temperature structure because this governs whether they are gravitationally unstable and thus capable of forming planets through gravitational instabilities \citep[e.g.,][]{Boss1997,Boley2009} or prone to luminosity outbursts \citep{Vorobyov2009}. In addition, knowledge of temperature is required to derive disk masses, and thus the amount of material available for planet formation, from continuum observations. Finally, the temperature sets the chemical composition of the planet-forming material, for example, through the sequential freeze-out of volatiles as the temperature decreases at larger distances from the central star \mbox{\citep[e.g.,][]{Oberg2011}}. 

The radial onset of CO freeze-out, the CO snowline, has been located in several protoplanetary disks, revealing that these disks have a large reservoir of cold ($\lesssim$25 K) gas with CO freeze-out starting at a few tens to $\sim$100 AU from the star \citep{Qi2013,Qi2015,Qi2019,Oberg2015,Dutrey2017}. In contrast to these mature disks, the young embedded disk in L1527 shows no signs of CO freeze-out \citep{vantHoff2018b} in agreement with model predictions \citep{Harsono2015}. But whether all young disks are warm and if so, when they start to cool, remain open questions. Moreover, while mature disks can be described with a power law midplane temperature profile, this is not necessarily the case for embedded disks as material may be heated in shocks at the disk-envelope interface (centrifugal barrier; e.g., \citealt{Sakai2014a}). Here, we study the gas temperature of the young disk-like structure around one component of the Class 0 protostellar system IRAS 16293--2422. 

IRAS 16293--2422 (hereafter IRAS 16293) is a well-studied deeply embedded protostellar binary in the L1689 region in $\rho$ Ophiuchus ($d \sim$140 pc; \citealt{Dzib2018}). The separation between the two components, often referred to as IRAS 16293A and IRAS 16293B, is $\sim$5.1$^{\prime\prime}$  ($\sim$715 AU; \citealt{Wootten1989,Mundy1992,Looney2000,Chandler2005,Pech2010}) and both protostars show compact millimeter continuum emission on 100 AU scales \citep{Looney2000,Schoier2004}. IRAS 16293A has a velocity gradient in the NE-SW direction that could be attributed to a rotating disk viewed edge-on, while rotating motion is hardly detected toward source B, suggesting a near face-on orientation \citep{Pineda2012,Favre2014,Oya2016}. The protostellar masses are estimated to be $M_A \sim 1 M_\odot$ and $M_B \sim 0.1 M_\odot$ \citep{Bottinelli2004,Caux2011}, but two  continuum sources ($\sim$0.38$^{\prime\prime}$ or $\sim$50 AU separation) have been detected toward IRAS 16293A at cm and mm wavelenghts, suggesting that IRAS 16293A itself may be a binary as well \citep{Wootten1989,Chandler2005,Pech2010}, or even a triple system if A1 is in fact a very tight binary \citep{Hernandez-Gomez2019}. 

Existing temperature structures for IRAS 16293 are based on modeling of continuum emission \citep[e.g.,][]{Schoier2002} and spatially unresolved observations of water and oxygen lines with the Infrared Space Observatory (ISO), assuming an infalling envelope around a single protostar \citep[e.g.,][]{Ceccarelli2000,Crimier2010}. The first detailed 3D modeling of the continuum, $^{13}$CO and C$^{18}$O emission, including two radiation sources ($\sim$18 $L_\odot$ for source A and $\sim$3 $L_\odot$ for source B), was presented by \citet{Jacobsen2018}. All these temperature profiles are consistent with water ice desorption ($\sim$100~K) at $\sim$100--150~AU.  

IRAS 16293 was the first low-mass protostar for which complex organics were detected \citep{vanDishoeck1995,Cazaux2003}. The ALMA Protostellar Interferometric Line Survey (PILS; an unbiased spectral survey between 329 and 363 GHz at 0.5$^{\prime\prime}$ resolution presented by \citealt{Jorgensen2016}) fully revealed the chemical richness of this source with approximately one line detected per 3 km s$^{-1}$. The large frequency range covered makes the PILS data also very well-suited to study the temperature structure through ratios of line emission from different transitions within one molecule. Particularly good tracers of temperature are H$_2$CO and H$_2$CS \citep[e.g.,][]{Mangum1993,vanDishoeck1993,vanDishoeck1995} for which multiple lines are covered by the PILS survey \citep{Persson2018}. 

H$_2$CO and H$_2$CS are slightly asymmetric rotor molecules and have their energy levels designated by the quantum numbers $J$, $K_{\rm{a}}$ and $K_{\rm{c}}$. Since transitions between energy levels with different $K_{\rm{a}}$ values operate only through collisional excitation, line ratios involving different $K_{\rm{a}}$ ladders are good tracers of the kinetic temperature \citep[e.g.,][]{Mangum1993,vanDishoeck1993,vanDishoeck1995}. Moreover, transitions from different $K_{\rm{a}}$ levels connecting the same $J$ levels are closely spaced in frequency such that they can be observed simultaneously. Therefore, line ratios from $K_{\rm{a}}$ transitions within the same $\Delta J$ transition can provide a measure of the kinetic temperature unaffected by relative pointing uncertainties, beam-size differences and absolute calibration uncertainties. 

In this paper we focus on H$_2$CS to derive a temperature profile for the disk-like structure around IRAS 16293A, because too few unblended lines were available for H$_2$CO and H$_2^{13}$CO, and the H$_2$CO and D$_2$CO lines are optically thick. The observations are briefly described in Sect.~\ref{sec:Observations}. In Sect.~\ref{sec:Results and analysis}, we present temperature maps based on H$_2$CS line ratios and rotation diagrams, showing that the temperature remains between $\sim$100 and $\sim$175 K out to $\sim$150 AU, consistent with envelope temperature profiles, but flattens in the inner $\sim$100 AU. At the current spatial resolution of the data, no jump in temperature is seen at the disk-envelope interface. This temperature profile is further discussed in Sect.~\ref{sec:Discussion} and the conclusions are summarized in Sect.~\ref{sec:Conclusions}. 

\begin{figure}
\centering
\includegraphics[width=0.5\textwidth,trim={0cm 15.3cm 7.7cm 1.5cm},clip]{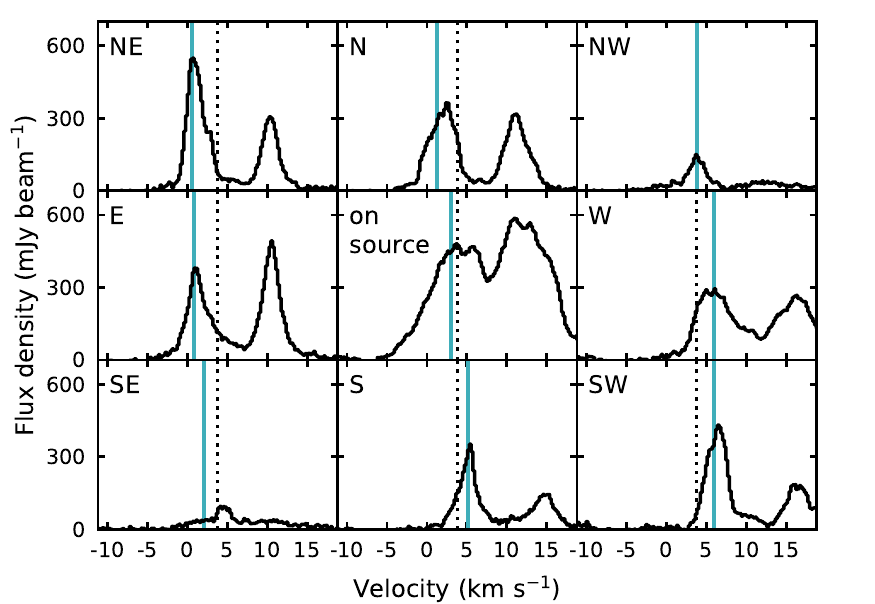}
\caption{Spectra of the H$_2$CS 10$_{0,10}$ -- 9$_{0,9}$ line at nine positions toward IRAS 16293A illustrating the variation in line blending across the source. The \textit{central panel} shows the spectrum in the central pixel (continuum peak), the other spectra are taken at a beam offset (0.5$\arcsec$) in the different directions as denoted in the top left corner of each panel. The vertical dotted line marks the system velocity of 3.8 km~s$^{-1}$, and the vertical blue solid line indicates the peak velocity from the CH$_3$OH line used to isolate the target lines (in this case H$_2$CS 10$_{0,10}$ -- 9$_{0,9}$). }
\label{fig:Spectra}
\end{figure}

\begin{figure}
\centering
\includegraphics[width=0.47\textwidth,trim={0.9cm 3.2cm 7.7cm 0.7cm},clip]{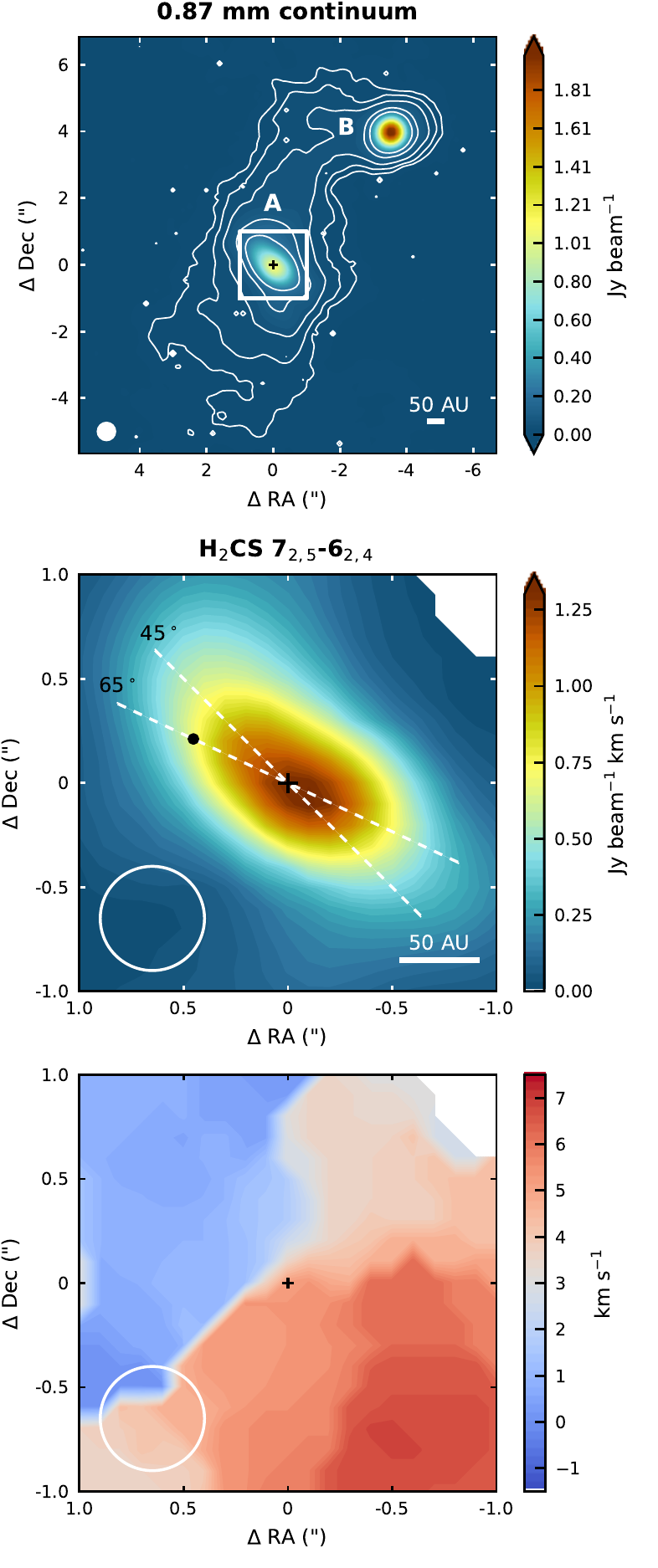}
\caption{Continuum image of IRAS 16293 at 0.87 mm (\textit{top panel}) with contour levels at 0.5\%, 1\%, 2\%, 5\% and 10\% of the peak flux. The white box marks the region around the A source for which the velocity-corrected integrated emission (VINE) map of the H$_2$CS 7$_{2,5}$ -- 6$_{2,4}$ line at 240.549 GHz ($E_{\rm{u}}$ = 99 K) is shown in the \textit{middle panel} and the peak velocity at each position in the \textit{bottom panel}. The continuum peak position of IRAS 16293A is marked with a cross in all panels and the beam is depicted in the lower left corners. The white lines in the \textit{middle panel} indicate the position angle of the major axis of the disk-like structure found by \citet{Oya2016} (65$^\circ$) and our by-eye best matching position angle of 45$^\circ$. The black circle marks the position where the spectra in Figure~\ref{fig:H2CS_pv} are extracted.}
\label{fig:VINEmap}
\end{figure}


\section{Observations} \label{sec:Observations}

The Protostellar Interferometric Line Survey (PILS) is an unbiased spectral survey of the low-mass protobinary IRAS 16293 with the Atacama Large Millimeter/submillimeter Array (ALMA) covering frequencies between 329.147 and 362.896 GHz in Band 7 (project-id: 2013.1.00278.S). Multiple lines from H$_2$CO, H$_2$CO isotopologues and H$_2$CS fall within this spectral range. $\Delta J$ transitions with multiple $K_{\rm{a}}$ transitions available are listed in Table~\ref{tab:Lineoverview}. The transition frequencies and other line data were taken from
the CDMS database \citep{Muller2001,Muller2005}. The H$_2$CO and entries are based on \citet{Bocquet1996}, \citet{Cornet1980}, \citet{Brunken2003}, and \citet{Muller2017}, and the H$_2^{13}$CO entries are taken from \citet{Muller2000}. Entries of the deuterated isotopologues are based on \citet{Dangoisse1978}, \citet{Bocquet1999}, and \citet{Zakharenko2015}. H$_2$CS data are provided by \citet{Fabricant1977}, \citet{Maeda2008}, and \citet{Muller2019}. 
 
The data have a spectral resolution of 0.2 km~s$^{-1}$ (244 kHz), a restoring beam of 0.5$\arcsec$ ($\sim$70 AU), and a sensitivity of 7--10 mJy beam$^{-1}$ channel$^{-1}$, or $\sim$5 mJy beam$^{-1}$ km s$^{-1}$ across the entire frequency range. The phase center is located between the two continuum sources at $\alpha$(J2000)~=~16$^{\rm{h}}$32$^{\rm{m}}$22$\fs$72; $\delta$(J2000)~=~$-$24$\degr$28$\arcmin$34$\farcs$3. In this work we focus on the A source ($\alpha$(J2000)~=~16$^{\rm{h}}$32$^{\rm{m}}$22$\fs$873; $\delta$(J2000)~=~
$-$24$\degr$28$\arcmin$36$\farcs$54). A detailed description of the data reduction and continuum subtraction can be found in \citet{Jorgensen2016}. 

The PILS program also contains eight selected windows in Band 6 ($\sim$230 GHz) at the same angular resolution as the Band 7 data (0.5$\arcsec$; project-id: 2012.1.00712.S). These data have a spectral resolution of 0.15 km~s$^{-1}$ (122 kHz) and a sensitivity of $\sim$4 mJy beam$^{-1}$ channel$^{-1}$, or $\sim$1.5 mJy beam$^{-1}$ km s$^{-1}$. The data reduction proceeded in the same manner as the Band 7 data \citep{Jorgensen2016}. Several H$_2$CS $J_{K_{\rm{a}}K_{\rm{c}}}$ = 7$_{xx}$ -- 6$_{xx}$ transitions are covered by this spectral setup (see Table~\ref{tab:Lineoverview}). 

IRAS 16293 was also observed at 0.5$\arcsec$ resolution by program 2016.1.01150.S (PI: Taquet). This dataset has a spectral resolution of 0.15 km~s$^{-1}$ (122 kHz) and a sensitivity of $\sim$1.3 mJy beam$^{-1}$ channel$^{-1}$, or $\sim$0.5 mJy beam$^{-1}$ km s$^{-1}$ and covers the H$_2$CS 7$_{1,7}$ -- 6$_{1,6}$ transition (Table~\ref{tab:Lineoverview}). The data reduction is described in \citet{Taquet2018}.


\section{Results} \label{sec:Results and analysis}

\subsection{Kinematics}

The incredible line-richness of IRAS 16293 (on average one line per 3.4 MHz) means that many lines are blended \citep[e.g.,][]{Jorgensen2016}. In addition, the large velocity gradient due to the near edge-on rotating structure around the A source \citep[\mbox{$\sim$6 km~s$^{-1}$;}][]{Pineda2012,Favre2014} results in varying line widths and hence varying degrees of blending at different positions (see Fig.~\ref{fig:Spectra}). Therefore, we make use of the method outlined in \citet{Calcutt2018} to isolate the formaldehyde and thioformaldehyde lines: the peak velocity in each pixel is determined using a bright methanol transition (7$_{3,5}$ -- 6$_{4,4}$ at 337.519 GHz) and this velocity map is then used to identify the target lines listed in Appendix~\ref{ap:Linelist}.   

Of the 18 formaldehyde and 15 thioformaldehyde lines, only the H$_2$CO 5$_{1,5}$ -- 4$_{1,4}$, H$_2^{13}$CO 5$_{2,4}$ -- 4$_{2,3}$, H$_2^{13}$CO 5$_{4,1}$ -- 4$_{4,0}$, D$_2$CO 6$_{3,4}$ -- 5$_{3,3}$ and H$_2$CS 7$_{2,5}$ -- 6$_{2,4}$ lines can be fully isolated. Figure~\ref{fig:VINEmap} shows the Velocity-corrected INtegrated emission (VINE) map for H$_2$CS 7$_{2,5}$ -- 6$_{2,4}$, that is, the moment zero map but integrated over different velocities in each pixel \citep{Calcutt2018}, as well as a map of the corresponding peak velocities. The emission shows an elongated structure in the northeast-southwest direction along the velocity gradient. The H$_2^{13}$CO 5$_{1,5}$ -- 4$_{1,4}$, H$_2^{13}$CO 5$_{2,3}$ -- 4$_{2,2}$, D$_2$CO 6$_{2,5}$ -- 5$_{2,4}$ and H$_2$CS 10$_{2,9}$ -- 9$_{2,8}$ lines are severely blended ($<$ 4 MHz to another line) and excluded from the analysis (see Table~\ref{tab:Lineoverview} for details). All other lines are blended to some degree. In most cases this means that one or both of the line wings in the central pixels ($\sim$0.5$\arcsec$ radius) overlap with the wing of another line.

\begin{figure*}
\centering
\includegraphics[width=.94\textwidth,trim={0cm 0.4cm 0cm 0.99cm},clip]{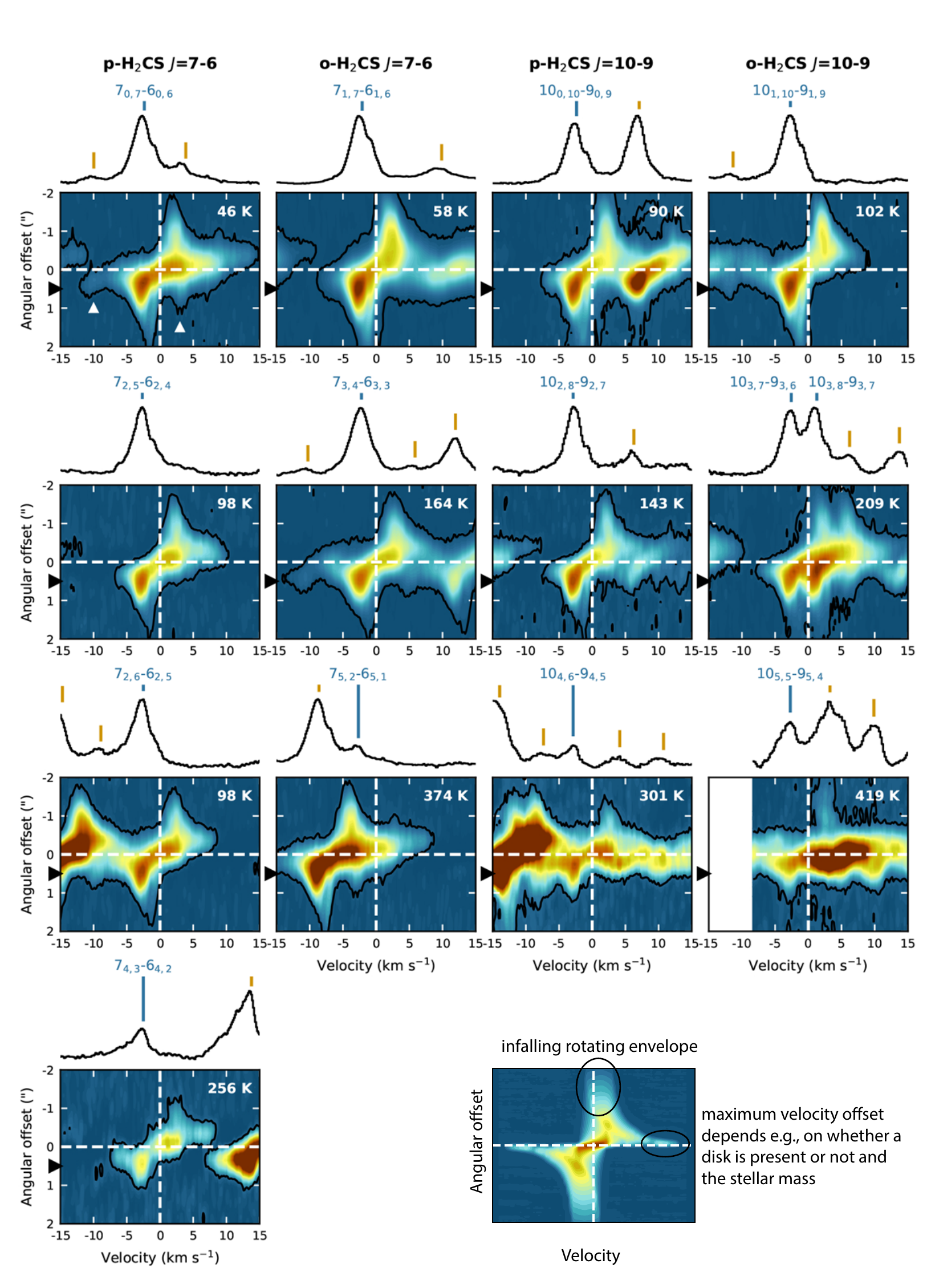}
\caption{Position-velocity diagrams of the H$_2$CS lines along the major axis (PA = 65$^\circ$) of the disk-like structure (positive angular offsets denote the northeast direction, i.e. blueshifted emission, negative offsets denote the southwest direction, i.e., redshifted emission). The intensity (color) scale is normalized to the brightest line in each panel. The black contour denotes the 3$\sigma$ contour. The dashed white lines mark the source position and systemic velocity of 3.8 km~s$^{-1}$ (shifted to 0 km s$^{-1}$). The upper level energy is denoted in the top right corner of each panel. The spectra at $\sim$0.5$\arcsec$ northeast of the source (indicated by black triangles left of the vertical axes) are shown on top of the pv-diagram panels. The H$_2$CS lines are identified by a vertical blue line, other lines by vertical orange lines. The white triangles in the top left panel highlight blending of the 7$_{0,7}$ -- 6$_{0,6}$ line. For reference, the bottom right panel shows the pv-diagram for a thin disk model with keplerian disk and infalling rotating envelope. }
\label{fig:H2CS_pv}
\end{figure*}

An overview of the 14 H$_2$CS lines is shown in Fig.~\ref{fig:H2CS_pv} in the form of position-velocity (pv) diagrams along the major axis of the disk-like structure (PA = 65$^\circ$; \citealt{Oya2016}) and spectra extracted $\sim$0.5$\arcsec$ to the northeast of the source center (see Fig.~\ref{fig:VINEmap}). Similar pv-diagrams for H$_2$CO, H$_2^{13}$CO and D$_2$CO are presented in Appendix~\ref{ap:pvdiagrams}. All lines show signs of rotation, that is, low velocities at large angular offsets and high velocities closer to the source center. H$_2$CO and H$_2$CS emission extends out to $\gtrsim$2$\arcsec$, while emission from the less abundant isotopologues H$_2^{13}$CO and D$_2$CO is more compact ($\lesssim$1$\arcsec$), probably due to sensitivity. The emission also becomes more compact for transitions with higher upper level energies, which are expected to trace the warmest inner region. The H$_2$CO lines show red-shifted absorption for velocities slightly above the systemic velocity of 3.8 km~s$^{-1}$. 

\begin{figure*}
\centering
\includegraphics[width=\textwidth,trim={0cm 11.2cm 0cm 1.4cm},clip]{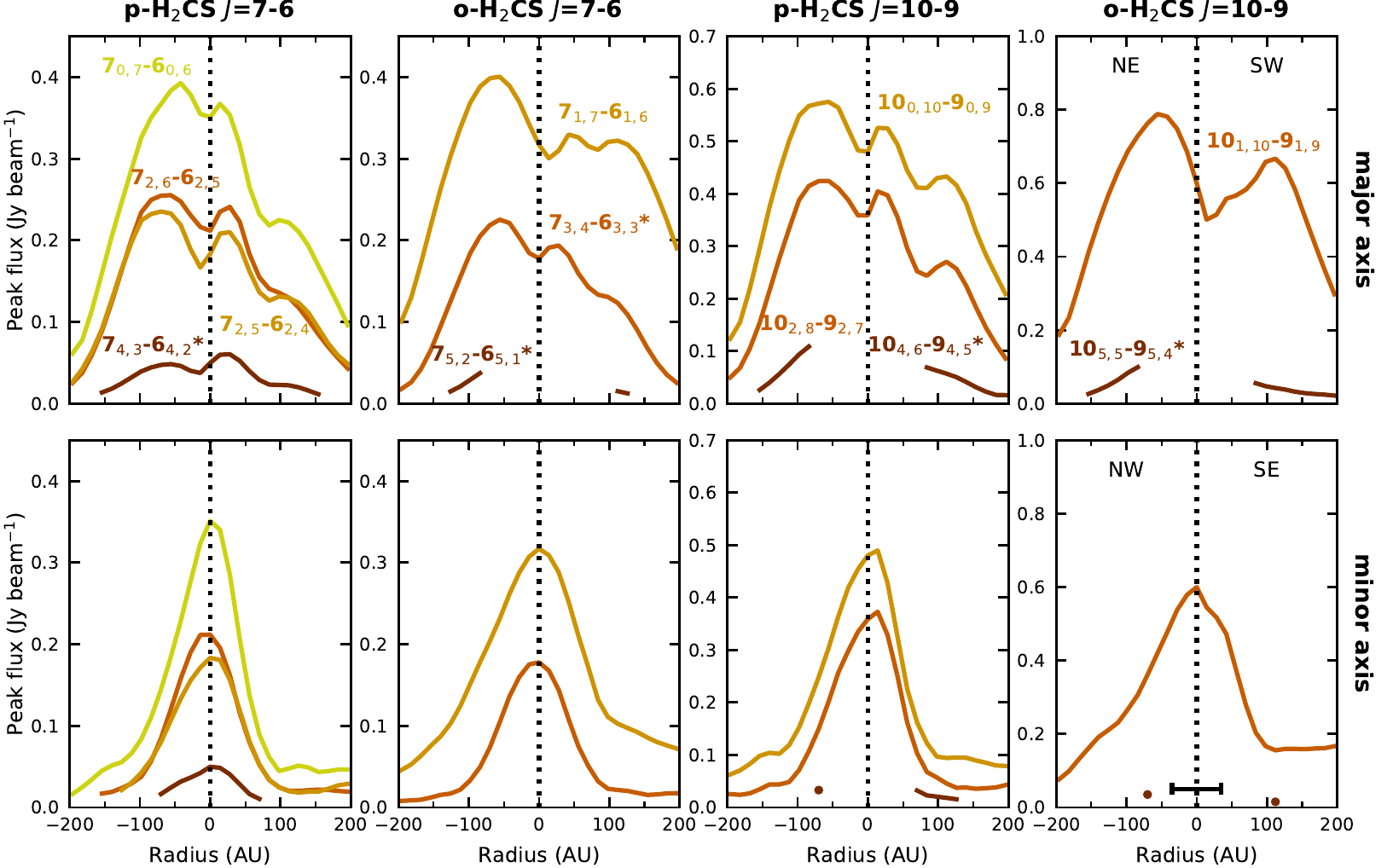}
\caption{Radial profiles of the H$_2$CS peak fluxes along the major axis (PA = 45$^\circ$; \textit{top panels}) and minor axis (PA = 135$^\circ$; \textit{bottom panels}) of the disk-like structure. In each panel, transitions with higher upper level energy are plotted in darker colors. Fluxes for transitions labeled with an asterisk ($\ast$) have been divided by 2 to account for two transitions with the same upper level energy and Einstein A coefficent at the same frequency (see Table~\ref{tab:Lineoverview}). The horizontal bar in the \textit{lower right panel} marks the beam size. The rms ranges between 2 and 8 mJy beam$^{-1}$.}
\label{fig:Peakflux_radial}
\end{figure*}

Based on large shifts in velocity from the systemic velocity (up to 14 km~s$^{-1}$) for the H$_2$CS 7$_{0,7}$ -- 6$_{0,6}$ line, \citet{Oya2016} suggested that a Keplerian disk may be present inside the centrifugal barrier (at $\sim$50 AU). However, careful analysis of the data shows that a weak line is present on both the blue and red side of this H$_2$CS line (see Fig.~\ref{fig:H2CS_pv}, top left panel). The large velocity gradient thus seems the result of line blending and not Keplerian rotation. This conclusion is reinforced by the velocity structure of the other H$_2$CS lines. The unblended 7$_{2,5}$ -- 6$_{2,4}$ line shows a maximum velocity offset of $\sim$6 km~s$^{-1}$ on the blue side, and $\sim$10 km~s$^{-1}$ on the red side. Unblended wings from the other lines show similar offsets. The unblended H$_2$CS velocity offsets are comparable to the ones for OCS $J = 19-18$ \citep{Oya2016}, except for OCS the offset is $\sim$8 km~s$^{-1}$ on the red side and $\sim$10 km~s$^{-1}$ on the blue side. \citet{Oya2016} showed that the OCS emission cannot be explained by Keplerian motions, but requires both rotation and infall. A Keplerian component in the inner part of the rotating-infalling disk-like structure can thus not be established with the current observations. 

\subsection{Peak fluxes} \label{sec:Peakflux} 

Most lines show some degree of blending, such that generally, in the central pixels ($\sim$0.5$\arcsec$ radius), one or both of the line wings overlap with the wing of another line. As this does not significantly influence the peak flux, we extract the peak flux per pixel in order to make spatial maps of the line ratios and hence temperature structure. Pixels with too much line blending to extract a reliable peak flux are excluded. Since the data cover only one ortho-H$_2$CO and one para-H$_2$CO line, the main isotopologue lines cannot be used for a temperature measurement and we exclude H$_2$CO from further analysis. The D$_2$CO 6$_{4,2}$ -- 5$_{4,1}$ and 6$_{4,3}$ -- 5$_{4,2}$ lines, as well as the H$_2$CS 10$_{3,7}$ -- 9$_{3,6}$ and 10$_{3,8}$ -- 9$_{3,7}$ lines, are located within 4 MHz of each other and are too blended to obtain individual peak fluxes (see also Figs.~\ref{fig:H2CS_pv} and ~\ref{fig:D2CO_pv}). Maps of the peak fluxes (moment 8 maps) of the remaining H$_2^{13}$CO, D$_2$CO and H$_2$CS lines without severe blending are presented in Appendix~\ref{ap:Peakflux} and radial profiles for the H$_2$CS lines are shown in Fig.~\ref{fig:Peakflux_radial}. 

All H$_2$CS lines show an elongated structure with the major axis at $\sim$45$^\circ$ (Figs.~\ref{fig:VINEmap} and \ref{fig:Peakflux_H2CS}). This is slightly different from the 65$^\circ\pm$10$^\circ$ derived by \citet{Oya2016}, but radial profiles along a position angle of 45$^\circ$ or 65$^\circ$ are very similar. For all lines, the emission peaks on either side of the continuum peak, as has been seen for most other species toward IRAS 16293A \citep[see e.g.,][]{Calcutt2018,Manigand2019}. This could correspond to the double peaks seen in cm continuum emission \citep{Wootten1989,Chandler2005,Pech2010}, although those positions are slightly off \citep[see discussion in][]{Calcutt2018}. Alternatively, they can be due to an edge-on rotating toroidal structure, as, for example, seen in synthetic CO images \citep{Jacobsen2018}. The NE peak is $\sim$60~AU offset from the continuum peak, except for the 7$_{0,7}$ -- 6$_{0,6}$ transition that peaks around $\sim$40~AU (see Fig.~\ref{fig:Peakflux_radial}). The SW peak is around $\sim$25~AU, and is weaker than the NE peak, except for the 7$_{4,3}$ -- 6$_{4,2}$ transition. Most H$_2$CS lines seem to have a weaker third peak $\sim$120~AU SW off the source center. The H$_2^{13}$CO and D$_2$CO lines also show an elongated double-peaked structure, albeit more compact and without the third peak seen for H$_2$CS (Fig.~\ref{fig:Peakflux_H2CO}). In addition, the northern part is more elongated perpendicular to the major axis than for the H$_2$CS lines. 

\begin{figure*}
\centering
\includegraphics[width=\textwidth,trim={0cm 14.3cm 0cm .9cm},clip]{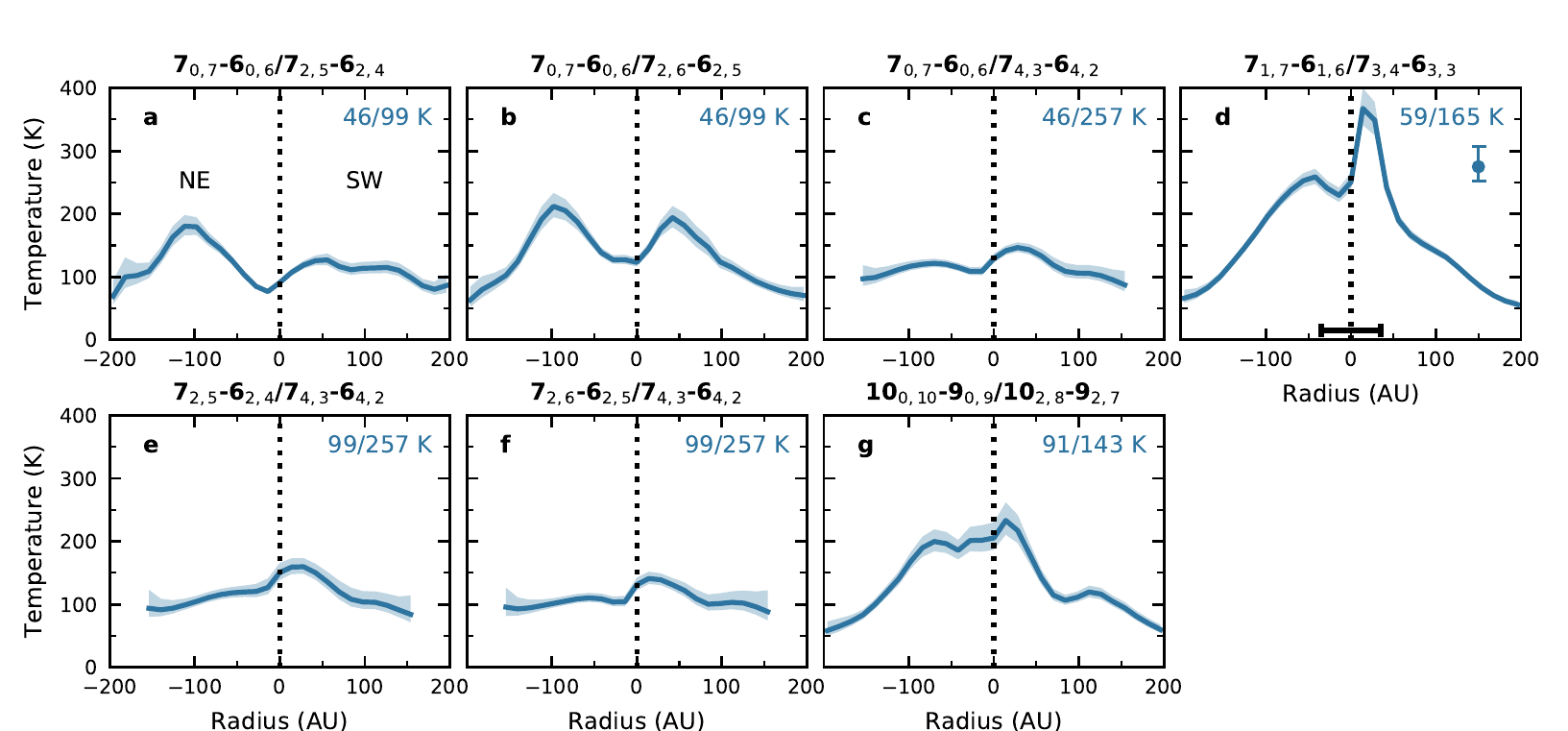}
\caption{Radial temperature profiles derived from H$_2$CS line ratios (listed above the panels) along the major axis of the disk-like structure (PA = 45$^\circ$) using the RADEX results. The upper level energies of the transitions are listed in the top right corner of each panel. The shaded area represents the 1$\sigma$ uncertainty based on the noise level of the observations. The vertical bar in the \textit{top right panel} shows a 3$\sigma$ error bar representative for the inner 100 AU and the horizontal bar marks the beam size.}
\label{fig:Tradial-major_H2CS}
\end{figure*}

\subsection{Temperature} \label{sec:Temperature} 

Only H$_2$CS lines are used in the temperature analysis, for the reasons outlined below. For H$_2^{13}$CO there are three line ratios that probe temperature (5$_{0,5}$ -- 4$_{0,4}$/5$_{2,4}$ -- 4$_{2,3}$, 5$_{0,5}$ -- 4$_{0,4}$/5$_{4,1}$ -- 4$_{4,0}$ and 5$_{2,4}$ -- 4$_{2,3}$/5$_{4,1}$ -- 4$_{4,0}$), but for the two highest energy lines, the continuum seems to be oversubtracted to varying degrees in different pixels. Reliable peak fluxes can therefore not be obtained for these two lines, leaving only a single suitable H$_2^{13}$CO line and H$_2^{13}$CO is therefore not used to derive the temperature structure. For D$_2$CO there are six suitable line ratios. However, all these ratios are near unity, suggesting that the emission is optically thick (see Appendix~\ref{ap:Peakflux}). Their brightness temperature is $\sim$30~K, much lower than expected for the inner region of an envelope, as also shown for other optically thick lines in the PILS data (e.g., CH$_2$DOH; \citealt{Jorgensen2018}), indicating that those lines probe colder foreground material. 

The maps of the H$_2$CS peak fluxes are used to calculate line ratios, and these are converted into temperature maps using a grid of RADEX non-LTE radiative transfer models \citep{vanderTak2007}. The molecular data are obtained from the Leiden Atomic and Molecular Database (LAMDA; \citealt{Schoier2005}, which contains collisional rate coefficients for H$_2$CS from \citealt{Wiesenfeld2013}). The line ratios as a function of temperature for different H$_2$ densities and H$_2$CS column densities are shown in Fig.~\ref{fig:RADEX}. For densities $\gtrsim$10$^6$--10$^7$ cm$^{-3}$ and column densities $\lesssim$10$^{15}$ cm$^{-2}$ the line ratios are independent of density and column density, respectively, and thus good tracers of the temperature. For a disk-like structure, densities are expected to be $> 10^7$ cm$^{-3}$, so the exact value adopted for the density does not influence the derived temperatures and we use a value of $10^{10}$ cm$^{-3}$. Toward IRAS 16293B, a H$_2$CS column density of $1.3 \times 10^{15}$ cm$^{-2}$ was found \citep{Drozdovskaya2018}. As will be discussed later in this section, based on the line ratios, the H$_2$CS column density toward IRAS 16293A cannot exceed more than a few times $10^{15}$ cm$^{-2}$. A column density of $10^{14}$ cm$^{-2}$ is adopted in the temperature calculations. All H$_2$CS lines are optically thin at this column density. At a column of $10^{15}$ cm$^{-2}$, the 7$_{1,7}$ -- 6$_{1,6}$ and 10$_{1,10}$ -- 9$_{1,9}$ transitions become optically thick (see Fig.~\ref{fig:tau_H2CS}), lowering the derived temperature by $\lesssim$10~K (see Fig.~\ref{fig:RADEX}). In addition to a density and column density, a line width of 2 km s$^{-1}$ is adopted to convert the observed line ratios to temperature. Error bars on the line ratios are calculated from the image rms, and converted into error bars on the temperature using the RADEX calculations (see Fig.~\ref{fig:RADEX}). The error bars therefore depend on the sensitivity of the line ratio to the temperature and the relative uncertainty on the peak flux. Since collisional rate coefficients are not available for all observed transitions, we also performed LTE calculations. 

\begin{figure*}
\centering
\includegraphics[width=0.8\textwidth,trim={0cm 15.7cm 5cm 1.3cm},clip]{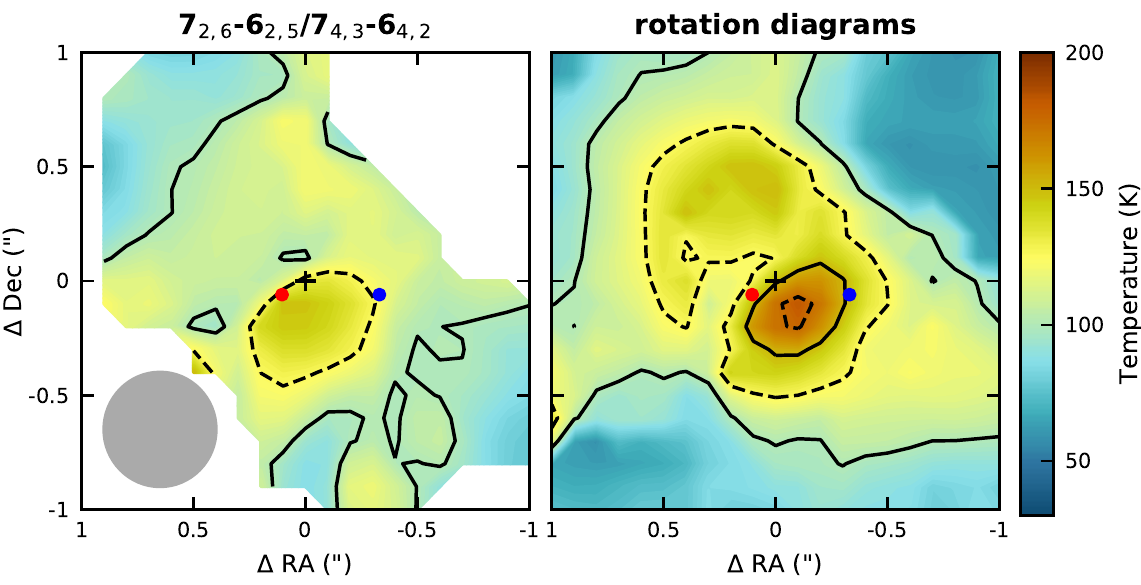}
\caption{Temperature in a $2^{\prime\prime} \times 2^{\prime\prime}$ (280 $\times$ 280 AU) region around IRAS 16293A derived from the H$_2$CS 7$_{2,6}$ -- 6$_{2,5}$/7$_{4,3}$ -- 6$_{4,2}$ line ratio (\textit{left panel}), and from the rotation diagrams in each pixel (\textit{right panel}). Solid contours indicate temperatures of 100 and 150 K and dashed contours are at 125 and 175 K. The continuum peak position is marked with a black cross. The positions of the unresolved binary components A1 and A2 in 2015 \citep{Hernandez-Gomez2019} are shown with a red and blue circle, resp. The beam is depicted in the lower left corner of the \textit{left panel}. } 
\label{fig:Tmaps_H2CS}
\end{figure*}

Radial temperature profiles along the major axis of the disk-like structure (PA = 45$^\circ$) are presented in Fig.~\ref{fig:Tradial-major_H2CS} for ratios with collisional rate coefficients and in Fig.~\ref{fig:Tradial-major_H2CS_LTE} for the remaining ratios. Figure~\ref{fig:Tradial_allmajor_H2CS} shows radial temperature profiles for different position angles. For clarity, we will refer to the panels in Fig.~\ref{fig:Tradial-major_H2CS} instead of to the quantum numbers of the line ratios in the following discussion. Along the major axis, all line ratios result in temperatures between $\sim$100 and $\sim$175 K out to $\sim$150 AU, with temperatures dropping to $\sim$75 K at $\sim$200 AU. The non-LTE RADEX results give temperatures higher by $\sim$20--50 K than the LTE results (see Fig.~\ref{fig:Tradial-major_H2CS_LTE}), suggesting that the lines are not completely thermalized. Panels \textit{d} and \textit{g} show temperatures $\gtrsim$150 K in the inner $\sim$100 AU. On the other hand, panels \textit{a} and \textit{b} display a decrease in temperature in the inner $\sim$100 AU (down to $\sim$100 K). The other three p-H$_2$CS ratios (panels \textit{c}, \textit{e} and \textit{f}) show a rather flat temperature profile, with temperatures of $\sim$100--150 K throughout the inner 150 AU. The five line ratios in Fig.~\ref{fig:Tradial-major_H2CS_LTE} cannot be used to constrain the temperature in the inner 100 AU due to severe line blending, but the temperature between $\sim$100--150 AU is $>$100 K. Another constraint can be provided by the ratio between the different $\Delta J$ transitions. The $10-9$ lines are brighter than the $7-6$ lines over the 200 AU radius studied, which means temperatures $\gtrsim$60~K (see Fig.~\ref{fig:RADEX}). 

The central decrease in panels \textit{a} and \textit{b} is not likely to be due to the continuum becoming optically thick, because a similar effect would then be expected in panels \textit{c}, \textit{e} and \textit{f}. Instead, it could be due to the fact that the 7$_{0,7}$ -- 6$_{0,6}$ line (which has the lowest upper level energy) peaks closer to the continuum position than the other lines, causing a central increase in the line ratio and hence decrease in the derived temperature. In addition, the temperature derived from these ratios is very sensitive to small changes in the fluxes, because the energy levels involved range from only 46 to 98 K (see Fig.~\ref{fig:RADEX}). The ratio changes from 1.4 to 1.5 if the temperature drops from 200 K to 150 K, and to 1.8 at 100 K. This means that a variation of $<$10\% in one of the peak fluxes can change the derived temperature from 200 K to 150 K, and a 20\% variation can change the temperature from 150 K to 100 K. Such variations can occur even though the transitions are present in the same dataset as can be seen from the difference between panels \textit{a} and \textit{b} (and Fig.~\ref{fig:Peakflux_radial}, upper left panel). The 7$_{2,6}$ -- 6$_{2,5}$ and 7$_{2,5}$ -- 6$_{2,4}$ transitions have the same upper level energy and Einstein A coefficient, and should thus be equally strong. However, in the SW there is a 100~K difference in the derived temperature using either of the two lines. This variation is slightly larger than the 3$\sigma$ uncertainty based on the rms noise, and may be due to blending with a weak line. 

The 7$_{4,3}$ -- 6$_{4,2}$ has a relatively high upper level energy (257~K) and therefore temperatures derived using this transition are less sensitive to flux changes. For example, as the temperature increases from 100 K to 200 K, the 7$_{0,7}$ -- 6$_{0,6}$/7$_{4,3}$ -- 6$_{4,2}$ ratio decreases from 11 to 4.1, and the 7$_{2,6}$ -- 6$_{2,5}$/7$_{4,3}$ -- 6$_{4,2}$ ratio decreases from 6.1 to 2.9. A temperature map for the $2^{\prime\prime} \times 2^{\prime\prime}$ region around IRAS 16293A derived from the 7$_{2,6}$ -- 6$_{2,5}$/7$_{4,3}$ -- 6$_{4,2}$ ratio (panel \textit{f}) is presented in Fig.~\ref{fig:Tmaps_H2CS}. The positions of the unresolved binary components of the A source, A1 and A2, are indicated as well. A1 and A2 have been observed to move relative to each other \citep[e.g.,][]{Loinard2002} with a global southwest movement. Since the observations used here are taken between 2014 and 2016, we plot the positions A1 and A2 had in 2015 \citep{Hernandez-Gomez2019}. Fig.~\ref{fig:Tmaps_H2CS} shows that the radial temperature profiles are not strongly affected by the chosen position angle or the binary components. 

The higher temperatures in panels \textit{d} and \textit{g} are likely due to optical depth effects. As can be seen from Fig.~\ref{fig:tau_H2CS}, the 7$_{1,7}$ -- 6$_{1,6}$ and 10$_{0,10}$ -- 9$_{0,9}$ transitions are among the first transitions to become optically thick. If these transitions are optically thick while the higher energy transitions remain optically thin, the flux of the lower energy transition is relatively too low and hence the derived temperature is too high. In turn, an estimate of the H$_2$CS column density can be made based on these results; for the 7$_{0,7}$ -- 6$_{0,6}$ transition to remain optically thin, the column density of para H$_2$CS cannot exceed more than a few times 10$^{15}$ cm$^{-2}$. 

\begin{figure*}
\centering
\includegraphics[width=\textwidth,trim={0cm 16.3cm 0cm 0.7cm},clip]{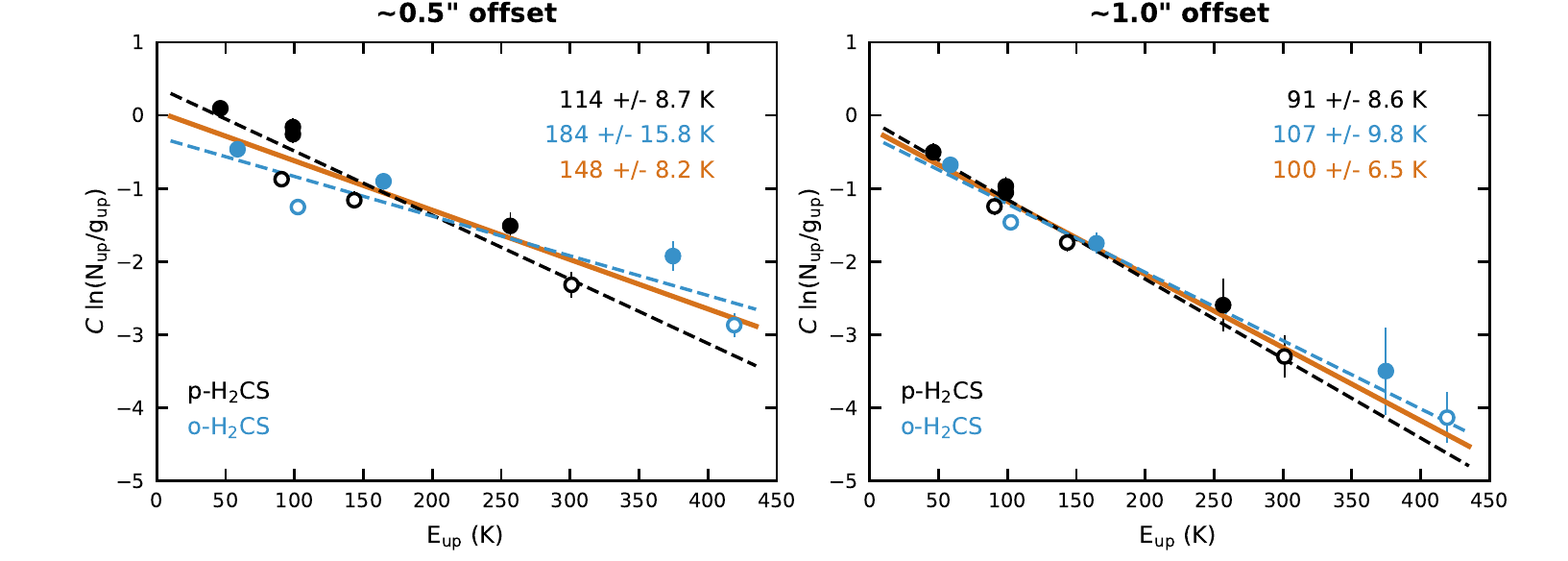}
\caption{Rotation diagrams for the H$_2$CS peak fluxes along the major axis (PA = 45$^\circ$) of the disk like structure at 0.5$^{\prime\prime}$ (\textit{left panel}) and 1.0$^{\prime\prime}$ (\textit{right panel}) from the source center. Para transitions are shown in black and ortho transitions in blue. An ortho-to-para ratio of 2 is assumed. $J=7-6$ transitions are displayed by filled symbols and $J=10-9$ transitions by open symbols. The error bars include a 10\% uncertainty. The best fit to all transitions is presented by the solid orange line, whereas the dashed lines show fits to only the ortho (blue) and only the para transitions (black). The corresponding rotation temperatures are listed in the top right corners.}
\label{fig:RotationDiagram}
\end{figure*}

To assess whether one of the H$_2$CS lines is affected by blending with a weak line at a similar frequency and therefore has its peak flux overestimated, rotation diagrams are made for every pixel in a $2^{\prime\prime} \times 2^{\prime\prime}$ region surrounding the continuum position. Representative diagrams at $0.5^{\prime\prime}$ and $1.0^{\prime\prime}$ offsets are presented in Fig.~\ref{fig:RotationDiagram}. An ortho-to-para ratio of 2 is assumed, but varying this between 1 and 3 changes the derived temperature by $\lesssim$30 K. At offsets $\gtrsim$0.8$^{\prime\prime}$ all transitions follow a straight line and similar temperatures are found when using all transitions, only ortho transitions or only para transitions in the fit. At smaller angular offsets the scatter becomes larger, but no transition is found to strongly deviate from the general trend. For the lower energy transitions ($E_{\rm{up}} <$ 200 K) this scatter is at least partly because some of the transitions become optically thick. 

\begin{figure*}
\centering
\includegraphics[width=\textwidth,trim={0cm 14.3cm 0cm .9cm},clip]{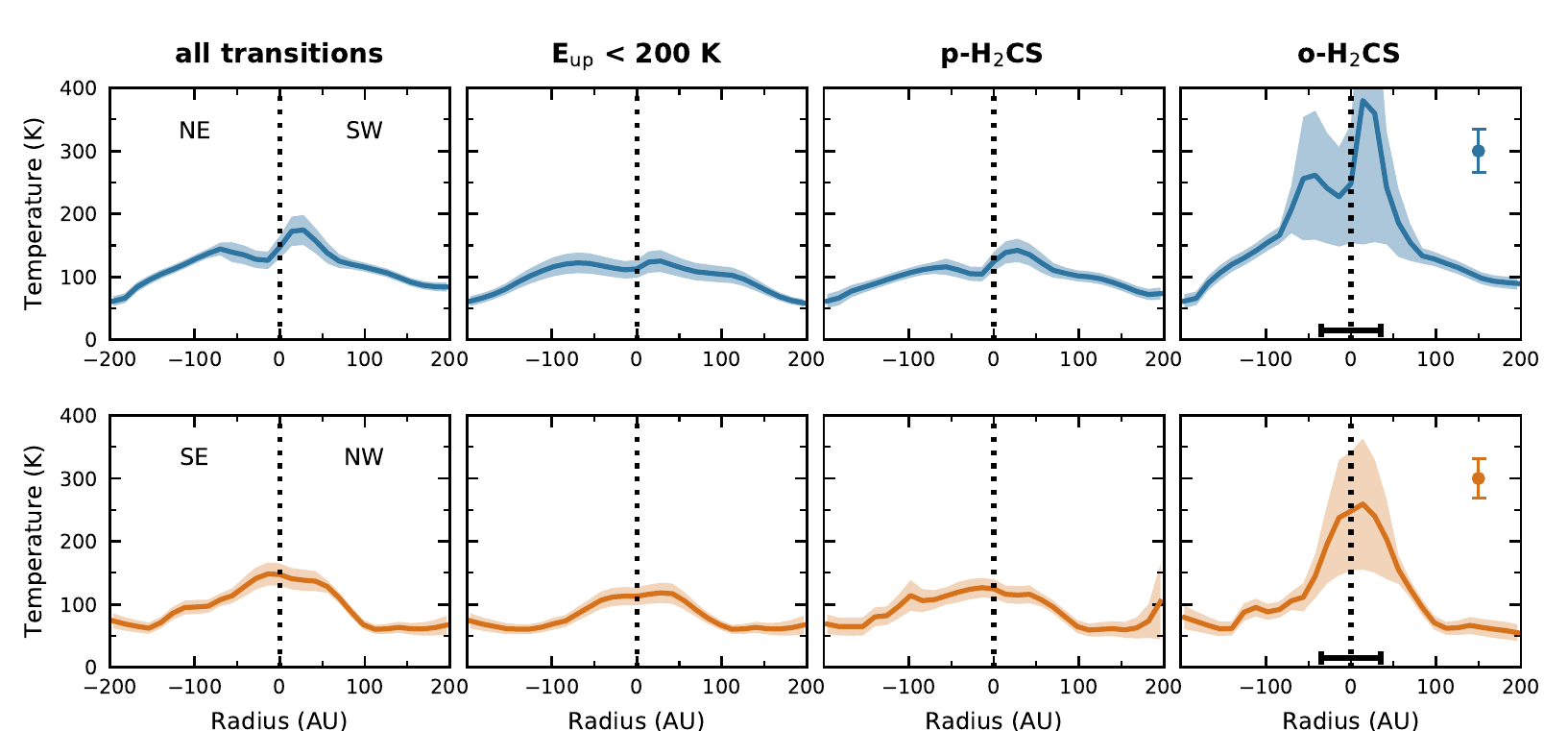}
\caption{Radial temperature profiles derived from the rotation diagrams including all transitions (\textit{first column}), only transitions with $E_{\rm{up}} <$ 200 K (\textit{second column}), only para transitions (\textit{third column}) or only ortho transitions (\textit{fourth column}). The \textit{top row} shows radial profiles along the major axis of the disk-like structure (PA = 45$^\circ$) and the \textit{bottom row} shows profiles along the minor axis. The shaded area represents the 1$\sigma$ uncertainty. The vertical bar in the \textit{right panels} shows a typical 3$\sigma$ error for the inner 100 AU and the horizontal bar marks the beam size. }
\label{fig:Tradial-RD_H2CS}
\end{figure*}

Radial profiles based on the temperatures derived from the rotation diagrams are shown in Fig.~\ref{fig:Tradial-RD_H2CS}. These profiles are constructed from fitting either all transitions, only transitions with $E_{\mathrm{up}} < 200$ K, only para transitions or only ortho transitions. The first three cases result in temperatures of $\sim$60 K at 200 AU rising up to $\sim$150 at source center. Using only the ortho transitions results in temperatures $> 200$ K in the inner 100 AU, likely because the lower energy transitions become optically thick. The central decrease in temperature seen for the 7$_{0,7}$ -- 6$_{0,6}$/7$_{2,5}$ -- 6$_{2,4}$ and 7$_{0,7}$ -- 6$_{0,6}$/7$_{2,6}$ -- 6$_{2,5}$ ratios (panels \textit{a} and \textit{b} in Fig.~\ref{fig:Tradial-major_H2CS}) is not seen in these profiles, and may be due to the 7$_{0,7}$ -- 6$_{0,6}$ transition peaking closer in and these ratios being not sensitive to temperature, as discussed above. 

\begin{figure*}
\centering
\includegraphics[width=.975\textwidth,trim={0cm 16.8cm 0cm .6cm},clip]{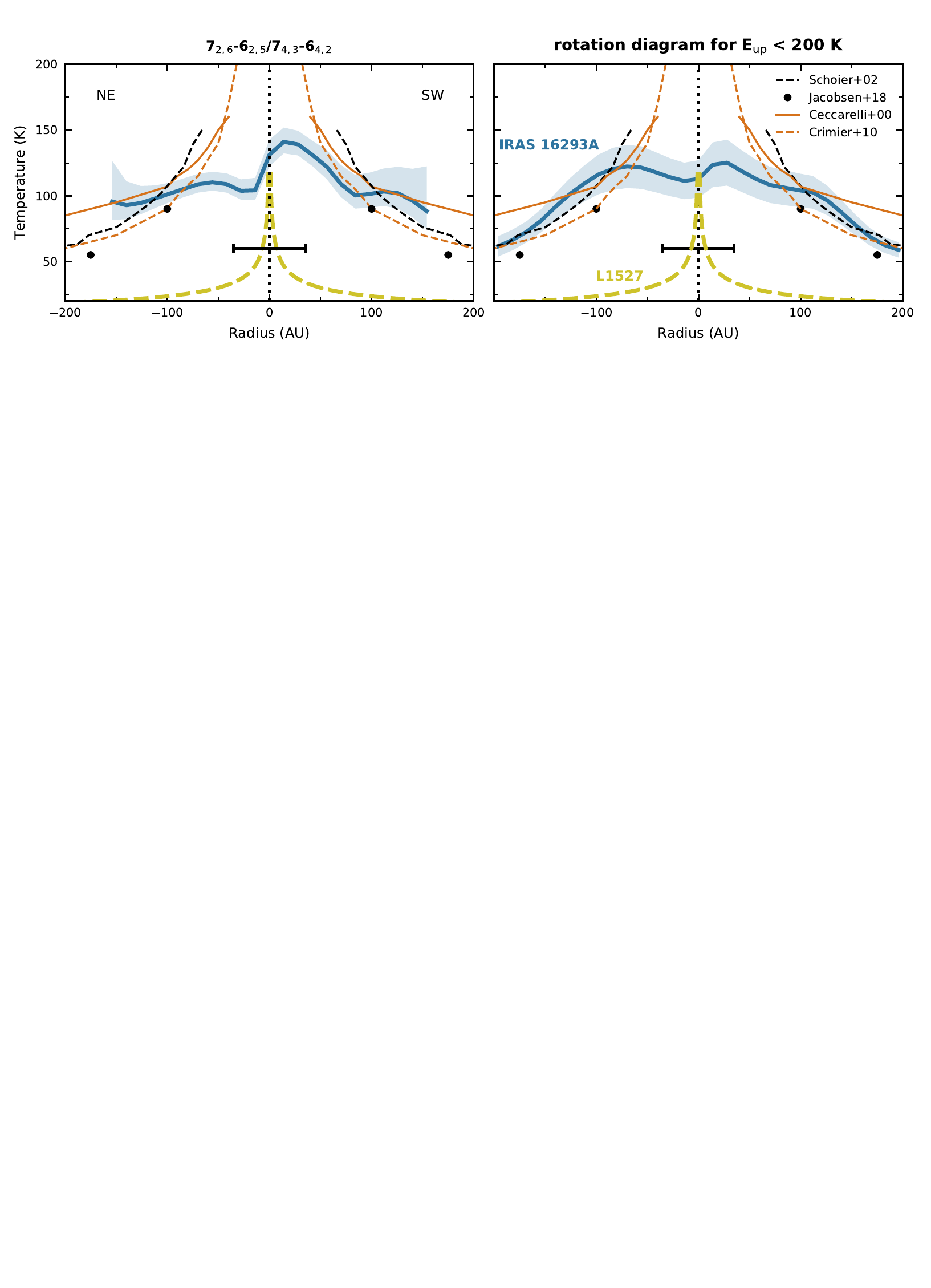}
\caption{Radial temperature profiles for IRAS 16293A derived in this work from H$_2$CS transitions (blue lines) compared to dust (black lines and circles) and gas (orange lines) temperature profiles from the literature. The H$_2$CS temperature profile in the \textit{left panel} is derived from the 7$_{2,6}$ -- 6$_{2,5}$/7$_{4,3}$ -- 6$_{4,2}$ line ratio and the profile in the \textit{right panel} is based on fits to the peak flux rotation diagrams including only transitions with $E_{\rm{up}} <$ 200 K. The blue shaded area represents the 1$\sigma$ uncertainty. The dashed black line shows the IRAS 16293 dust temperature assuming an envelope around around a central source of 27 $L_\odot$ \citep{Schoier2002}, and the black dots represents dust temperature contours for IRAS 16293A from detailed modeling including both protostars and the bridge structure \citep{Jacobsen2018}. The gas temperatures are based on H$_2$O observations with the ISO satellite using a model that computes the gas temperature through an envelope by solving a chemical network and equating the heating and cooling rates \citep{Ceccarelli1996}. The solid orange line shows the initial results assuming a single 27 $L_\odot$ protostar at 160 pc \citep{Ceccarelli2000}, the dashed orange line presents the revised results for a 22 $L_\odot$ protostar at 120 pc \citep{Crimier2010}. The dashed yellow line shows the midplane gas temperature for the Class 0/I disk L1527 based on optically thick $^{13}$CO and C$^{18}$O emission \citep{vantHoff2018b}. The horizontal bar marks the beam size of the H$_2$CS observations.}
\label{fig:Tprofiles}
\end{figure*}


\section{Discussion} \label{sec:Discussion}

Based on line ratios of several $J = 7-6$ and $J = 10-9$ H$_2$CS transitions, the temperature in the inner $\sim$ 150 AU of IRAS 16293A is $\sim$100--175 K, and drops to $\sim$75 K at $\sim$200 AU. In Fig.~\ref{fig:Tprofiles}, the H$_2$CS temperature profiles are compared to temperature profiles obtained from continuum radiative transfer modeling and from spatially unresolved H$_2$O observations with the ISO satellite \citep{Ceccarelli2000,Schoier2002,Crimier2010}. These latter profiles were derived assuming a spherically symmetric collapsing envelope structure around a single protostar. Recently, \citet{Jacobsen2018} performed detailed 3D radiative transfer modeling including two protostars. All previously derived temperatures show a very similar structure with a radial power law dependence of approximately $T \propto r^{-0.7}$ in the inner few hundred AU and a temperature of 100 K around $\sim$100 AU. The H$_2$CS temperature is consistent with the envelope temperature profiles down to radii of $\sim$100 AU, but lacks the temperature rise at smaller radii. 

Such a flattening of the temperature could be the result of efficient cooling by water molecules that are released into the gas phase at $\sim$100 K. However, water can also heat the gas through absorption of NIR photons from warm dust. \citet{Ceccarelli1996} studied the gas and dust temperature profile of an infalling envelope with a model that includes dynamics, chemistry, heating and cooling, and concluded that the gas and dust temperatures are generally well-coupled. For different model parameters (e.g., luminosity, mass accretion rate, stellar mass), the difference between the gas and dust temperature ranges from $\sim$10~K to more than a 100 K, but applying this model to IRAS 16293 (assuming a collapsing envelope around a single central source of 27 $L_\odot$) results in differences of only a few Kelvin between the gas and dust \citep{Ceccarelli2000,Crimier2010}. However, this model does not include a disk or disk-like structure. The presence of a disk may result in lower temperatures than in the case of only an envelope, as for example shown by the models from \citet{Whitney2003}, observations toward VLA~1623 \citep{Murillo2015} and analysis of H$_2^{18}$O observations toward four protostars in Perseus \citep{Persson2016}. 

A chemical depletion of H$_2$CS in the inner region is not likely to cause a flat instead of increasing temperature profile, because the line ratios are independent of abundance. The dust opacity could play a role in veiling the hottest inner region, but no absorption, as expected for colder gas in front of warmer dust, is observed. However, the situation is likely complicated with the dust and strongest lines being only marginally optically thick on scales of the beam, but much thicker on smaller unresolved scales. This may lead to quenching of the higher excitation lines while the emission from the lower excitation lines is only slightly reduced, and therefore result in a lower derived excitation temperature. On the other hand, this effect may be mitigated by the fact that multiple lines of sight are integrated in the beam. Detailed radiative transfer modeling is thus required to properly constrain the influence of the dust.

Figure~\ref{fig:Tprofiles} also shows the gas temperature profile for the embedded Class 0/I disk L1527 based on optically thick $^{13}$CO and C$^{18}$O observations \citep{vantHoff2018b}, illustrating that the temperature in the disk is $\sim$75~K lower than in the IRAS 16293A disk-like structure. This is consistent with the higher accretion rate and higher luminosity of IRAS 16293A ($\sim4\times10^{-6} - 5\times10^{-5} M_\odot$ yr$^{-1}$ and 18 $L_\odot$; \citealt{Schoier2002,Jacobsen2018}) compared to L1527 ($3\times10^{-7} M_\odot$ yr$^{-1}$ and 1.9-2.6 $L_\odot$; \citealt{Kristensen2012,Tobin2012,Tobin2013}). As the accretion rate drops in the late stages of the evolution of embedded protostellars, this suggests that the temperature of the disk or disk-like structure drops accordingly.

\citet{Oya2016} also used the 7$_{0,7}$ -- 6$_{0,6}$/7$_{2,5}$ -- 6$_{2,4}$ and 7$_{0,7}$ -- 6$_{0,6}$/7$_{4,3}$ -- 6$_{4,2}$ ratios (Fig.~\ref{fig:Tradial-major_H2CS}, panels \textit{a} and \textit{c}) to measure the gas temperature in the envelope, disk and envelope-disk interface (centrifugal barrier). Using the flux within a 0.5$^{\prime\prime}$ region at 0$^{\prime\prime}$, 0.5$^{\prime\prime}$ (70 AU) and 1.0$^{\prime\prime}$ (140 AU) offsets from the continuum peak position along the disk major axis (PA = 65$^\circ$) integrated over different velocity ranges at each position, they derive temperatures between 70 and 190~K. In addition, they find the highest temperatures at the 0.5$^{\prime\prime}$ offset position and attribute these temperature increases to weak accretion shocks at the centrifugal barrier ($\sim$50 AU) or to directly heating of a more extended envelope in front of the centrifugal barrier. Although we find a similar temperature range for the inner $\sim$150 AU, we do not see temperature peaks in the 7$_{0,7}$ -- 6$_{0,6}$/7$_{4,3}$ -- 6$_{4,2}$ (Fig.~\ref{fig:Tradial-major_H2CS}, panel \textit{c}). Peaks are visible for the 7$_{0,7}$ -- 6$_{0,6}$/7$_{2,5}$ -- 6$_{2,4}$ and 7$_{0,7}$ -- 6$_{0,6}$/7$_{2,6}$ -- 6$_{2,5}$ (Fig.~\ref{fig:Tradial-major_H2CS}, panels \textit{a} and \textit{b}), but this is likely due to the 7$_{0,7}$ -- 6$_{0,6}$ transition peaking slightly closer to the protostar than the other transitions in combination with the high sensitivity of these ratios to small changes in flux, as discussed in Sect.~\ref{sec:Temperature}. Moreover, no temperature peaks are found from the 7$_{2,5}$ -- 6$_{2,4}$/7$_{4,3}$ -- 6$_{4,2}$ and 7$_{2,6}$ -- 6$_{2,5}$/7$_{4,3}$ -- 6$_{4,2}$ ratios (Fig.~\ref{fig:Tradial-major_H2CS}, panels \textit{e} and \textit{f}) or from the rotational diagrams. Additional, higher resolution observations thus seem necessary to substantiate local temperature enhancements at the centrifugal barrier. 


\section{Conclusions}  \label{sec:Conclusions}

We used H$_2$CS line ratios observed in the ALMA-PILS survey to study the gas temperature in the disk-like structure around the Class 0 protostar IRAS 16293A. Because of the line-richness of this source, almost all H$_2$CS lines are blended in their line wings. Therefore, peak fluxes extracted per pixel are used in the analysis instead of integrated fluxes. Formaldehyde is not used because the optically thick H$_2$CO and D$_2$CO lines also trace colder foreground material, and not enough unblended H$_2^{13}$CO transitions were available. Our conclusions can be summarized as follows: 

\begin{itemize}
\item Strong evidence for a rotationally supported disk is still lacking: the high velocity wings of the H$_2$CS 7$_{0,7}$ -- 6$_{0,6}$ transition are the result of blending with nearby weak lines instead of a Keplerian disk. 

\item The temperature is between $\sim$100--175 K in the inner $\sim$150~AU, and drops to $\sim$75 K at $\sim$200~AU. This profile is consistent with envelope temperature profiles constrained on 1000 AU scales down to scales of $\sim$100 AU ($T \propto r^{-0.7}$ in the region where the dust becomes optically thick). However, the envelope profile breaks down in the inner most region that can now be probed by ALMA, because the observed temperature does not show a steep rise at radii $\lesssim$100 AU.   

\item The temperature in the disk-like structure around IRAS 16293A is $\sim$75 K higher than in the young Class 0/I disk L1527, on similar scales. 
\end{itemize}

\noindent The flattening of the temperature profile at radii $\lesssim$100 AU may be the result of efficient cooling by water that has sublimated from the grains in this region. Alternatively, the presence of a disk could affect the temperature structure, by having a larger fraction of the mass at higher density and lower temperatures. High resolution observations of optically thin lines are required to establish the details of the temperature profile in the inner $\sim$100 AU. 


\begin{acknowledgements} 
We would like to thank the referee for a prompt and positive report and Yuri Aikawa, Karin {\"O}berg, Jonathan Williams and Niels Ligterink for comments on the manuscript. Astrochemistry in Leiden is supported by the Netherlands Research School for Astronomy (NOVA). M.L.R.H acknowledges support from a Huygens fellowship from Leiden University. The research of J.K.J is supported by the European Union through the ERC Consolidator Grant ``S4F'' (grant agreement No 646908). This paper makes use of the following ALMA data: ADS/JAO.ALMA\#2012.1.00712.S, ADS/JAO.ALMA\#2013.1.00278.S and ADS/JAO.ALMA\#2016.1.01150.S. ALMA is a partnership of ESO (representing its member states), NSF (USA) and NINS (Japan), together with NRC (Canada), MOST and ASIAA (Taiwan), and KASI (Republic of Korea), in cooperation with the Republic of Chile. The Joint ALMA Observatory is operated by ESO, AUI/NRAO and NAOJ. 
\end{acknowledgements}


\bibliographystyle{aa} 
\bibliography{References} 


\begin{appendix}


\section{Formaldehyde and thioformaldehyde lines} \label{ap:Linelist}

Table~\ref{tab:Lineoverview} provides an overview of the H$_2$CO, H$_2^{13}$CO, D$_2$CO and H$_2$CS lines that have $\Delta J$ transitions with multiple $K_{\rm{a}}$ transitions in the spectral coverage of the ALMA-PILS survey. 

\begin{table*}[!h]
\addtolength{\tabcolsep}{-2pt}
\caption{Line list of formaldehyde and thioformaldehyde lines.}
\label{tab:Lineoverview}
\centering
\begin{tabular}{lcccccccl} 
\hline\hline
\\[-.3cm]
Species & Transition & Spin & Frequency & log($A_{ul}$) & $E_{u}$ & Used in & Dataset & Notes \\
        &           & isomer & (GHz)     & (s$^{-1}$) & (K) & analysis &  \\
\hline
\\[-.3cm]
H$_2$CO      & 5$_{0,5}$ -- 4$_{0,4}$ & para & 362.736 & -2.863 & 52 & - & 2013.1.00278.S & \\
             & 5$_{1,5}$ -- 4$_{1,4}$ & ortho &351.769 & -2.920 & 62 & - & 2013.1.00278.S &\\
\hline
\\[-.3cm]
H$_2^{13}$CO & 5$_{0,5}$ -- 4$_{0,4}$	& para & 353.812 &	-2.895 & 51	& - & 2013.1.00278.S &\\
             & 5$_{1,5}$ -- 4$_{1,4}$ & ortho & 343.326 & -2.952 & 61	& - & 2013.1.00278.S & Blended with H$_2$CS 10$_{2,9}$ -- 9$_{2,8}$ \\
             & 5$_{2,3}$ -- 4$_{2,2}$ & para & 356.176 & -2.962 & 99	& - & 2013.1.00278.S & CH$_2$DOH at 356.176 GHz \\
             & 5$_{2,4}$ -- 4$_{2,3}$ & para & 354.899 & -2.966 & 98	& - & 2013.1.00278.S & \\
             & 5$_{3,2}$ -- 4$_{3,1}$ & ortho	& 355.203 &	-3.083 &	158	& - & 2013.1.00278.S & \\
             & 5$_{3,3}$ -- 4$_{3,2}$ & ortho	& 355.191 &	-3.083 &	158	& - & 2013.1.00278.S & \\
             & 5$_{4,1}$ -- 4$_{4,0}$ & para	& 355.029 &	-3.334 &	240	& - & 2013.1.00278.S & At same frequency as 5$_{4,2}$ -- 4$_{4,1}$ \\
             & 5$_{4,2}$ -- 4$_{4,1}$ & para	& 355.029 &	-3.334 &	240	& - & 2013.1.00278.S & At same frequency as 5$_{4,1}$ -- 4$_{4,0}$ \\		
\hline  
\\[-.3cm]       
D$_2$CO & 6$_{0,6}$ -- 5$_{0,5}$	& para & 342.522 &	-2.933 & 	58 &	 - & 2013.1.00278.S & CH$_2$DOH at 342.522 GHz\\
        & 6$_{1,6}$ -- 5$_{1,5}$	& ortho & 330.674 &	-2.989 &		61 &	 - & 2013.1.00278.S & \\
        & 6$_{2,4}$ -- 5$_{2,3}$	& para & 357.871 &	-2.925 &		81 &	 - & 2013.1.00278.S & \\
        & 6$_{2,5}$ -- 5$_{2,4}$	& para & 349.631 &	-2.955 &		80 &	 - & 2013.1.00278.S & CH$_2$DOH at 349.635 GHz \\ 
        & 6$_{3,3}$ -- 5$_{3,2}$	& ortho & 352.244 &	-3.019 &		108 & - & 2013.1.00278.S & \\
        & 6$_{3,4}$ -- 5$_{3,3}$	& ortho & 351.894 &	-3.020 &		108 & - & 2013.1.00278.S & \\
        & 6$_{4,2}$ -- 5$_{4,1}$	& para & 351.492 &  -3.152 &	145 & - & 2013.1.00278.S & \\        
        & 6$_{4,3}$ -- 5$_{4,2}$	& para & 351.487 & -3.152 &		145 & - & 2013.1.00278.S & \\
        & 6$_{5,1}$ -- 5$_{5,0}$	& ortho & 351.196 &	-3.413 &		194	& - & 2013.1.00278.S & At same frequency as 6$_{5,2}$ -- 5$_{5,1}$  \\
        & 6$_{5,2}$ -- 5$_{5,1}$	& ortho & 351.196 &	-3.413 &	194	& - & 2013.1.00278.S & At same frequency as 6$_{5,1}$ -- 5$_{5,0}$  \\ 
\hline
\\[-.3cm]
H$_2$CS & 7$_{0,7}$ -- 6$_{0,6}$  & para & 240.267 & -3.688 & 46 & yes & 2012.1.00712.S & \\
        & 7$_{1,7}$ -- 6$_{1,6}$  & ortho & 236.727 & -3.717 & 59 & yes & 2016.1.01150.S & \\
        & 7$_{2,5}$ -- 6$_{2,4}$  & para & 240.549 & -3.724 & 99 & yes & 2012.1.00712.S & \\
        & 7$_{2,6}$ -- 6$_{2,5}$  & para & 240.382 & -3.725 & 99 & yes & 2012.1.00712.S & \\ 
        & 7$_{3,4}$ -- 6$_{3,3}$  & ortho & 240.394 & -3.776 & 165 & yes & 2012.1.00712.S & At same frequency as 7$_{3,5}$ -- 6$_{3,4}$ \\
        & 7$_{3,5}$ -- 6$_{3,4}$  & ortho & 240.393 & -3.776 & 165 & - & 2012.1.00712.S & At same frequency as 7$_{3,4}$ -- 6$_{3,3}$ \\
        & 7$_{4,3}$ -- 6$_{4,2}$  & para & 240.332 & -3.860 & 257 & yes & 2012.1.00712.S & At same frequency as 7$_{4,4}$ -- 6$_{4,3}$ \\
        & 7$_{4,4}$ -- 6$_{4,3}$  & para & 240.332 & -3.860 & 257 & - & 2012.1.00712.S & At same frequency as 7$_{4,3}$ -- 6$_{4,2}$ \\
        & 7$_{5,2}$ -- 6$_{5,1}$  & ortho & 240.262 & -3.998 & 375 & yes & 2012.1.00712.S & At same frequency as 7$_{5,3}$ -- 6$_{5,2}$ \\
        & 7$_{5,3}$ -- 6$_{5,2}$  & ortho & 240.262 & -3.998 & 375 & - & 2012.1.00712.S & At same frequency as 7$_{5,2}$ -- 6$_{5,1}$ \\
        & 10$_{0,10}$ -- 9$_{0,9}$ & para &	342.946 & -3.216 & 91  & yes & 2013.1.00278.S & \\
        & 10$_{1,9}$ -- 9$_{1,8}$	 & ortho & 348.534 & -3.199 & 105 & - & 2013.1.00278.S & Several lines at similar frequency \\
        & 10$_{1,10}$ -- 9$_{1,9}$ & ortho &	338.083 & -3.239	 & 102 & yes & 2013.1.00278.S & \\
        & 10$_{2,8}$ -- 9$_{2,7}$	 & para & 343.813 & -3.231	 & 143 & yes & 2013.1.00278.S & \\
        & 10$_{2,9}$ -- 9$_{2,8}$	 & para & 343.322 & -3.232	 & 143 & - & 2013.1.00278.S & Blended with H$_2^{13}$CO 5$_{1,5}$ -- 4$_{1,4}$\\
        & 10$_{3,7}$ -- 9$_{3,6}$	& ortho &  343.414 & -3.255  & 209 & - & 2013.1.00278.S & Blended with H$_2$CS 10$_{3,8}$ -- 9$_{3,7}$ \\
        & 10$_{3,8}$ -- 9$_{3,7}$	& ortho &  343.410 & -3.255	 & 209 & - & 2013.1.00278.S & Blended with H$_2$CS 10$_{3,7}$ -- 9$_{3,6}$ \\
        & 10$_{4,6}$ -- 9$_{4,5}$	& para &  343.310 & -3.290	 & 301 & yes & 2013.1.00278.S & At same frequency as 10$_{4,7}$ -- 9$_{4,6}$ \\
        & 10$_{4,7}$ -- 9$_{4,6}$	& para & 343.310 & -3.290	 & 301 & - & 2013.1.00278.S & At same frequency as 10$_{4,6}$ -- 9$_{4,5}$ \\
        & 10$_{5,5}$ -- 9$_{5,4}$	& ortho &  343.203 & -3.340	 & 419  & - & 2013.1.00278.S & Not detected \\
        & 10$_{5,6}$ -- 9$_{5,5}$	& ortho &  343.203 & -3.340	 & 419  & - & 2013.1.00278.S & Not detected \\
\hline       
\end{tabular}
\addtolength{\tabcolsep}{+2pt}
\end{table*}


\section{Position-velocity diagrams} \label{ap:pvdiagrams}

\begin{figure}[!h]
\centering
\includegraphics[trim={0cm 16cm 0cm 0.8cm},clip]{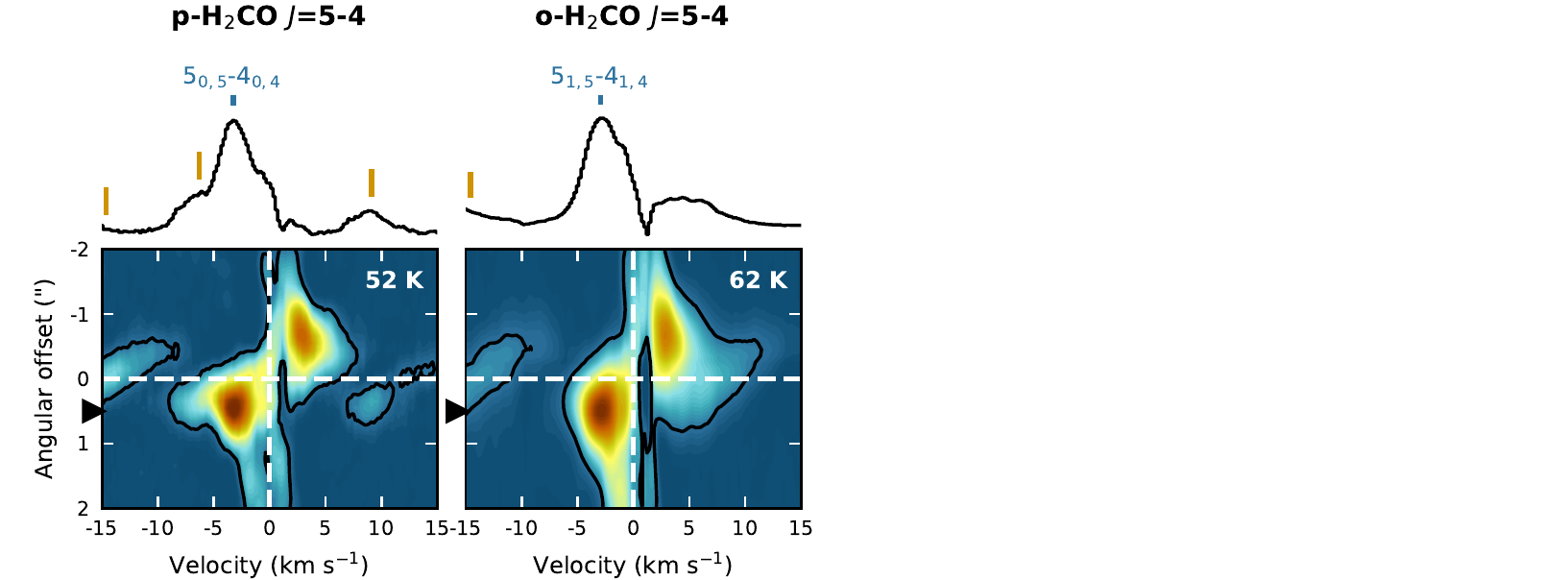}
\caption{Position-velocity diagrams of the H$_2$CO lines along the major axis of the disk-like structure (positive angular offsets denote the northeast direction, i.e., blueshifted emission, and negative offsets denote the southwest direction, i.e., redshifted emission). The intensity (color) scale is normalized to the brightest line in each panel. The black contour denotes the 3$\sigma$ contour. The dashed white lines mark the source position and systemic velocity of 3.8 km~s$^{-1}$ (shifted to 0 km~s$^{-1}$). The upper level energy is denoted in the top right corner of each panel. The spectra at $\sim$ 0.5$\arcsec$ northeast of the source (indicated by black  triangles left of the vertical axes) are shown on top of the pv-diagram panels. The H$_2$CO lines are identified by a vertical blue line, other lines by vertical orange lines. }
\label{fig:H2CO_pv}
\end{figure}

\begin{figure}[!h]
\centering
\includegraphics[trim={0cm 5.5cm 0cm 0.6cm},clip]{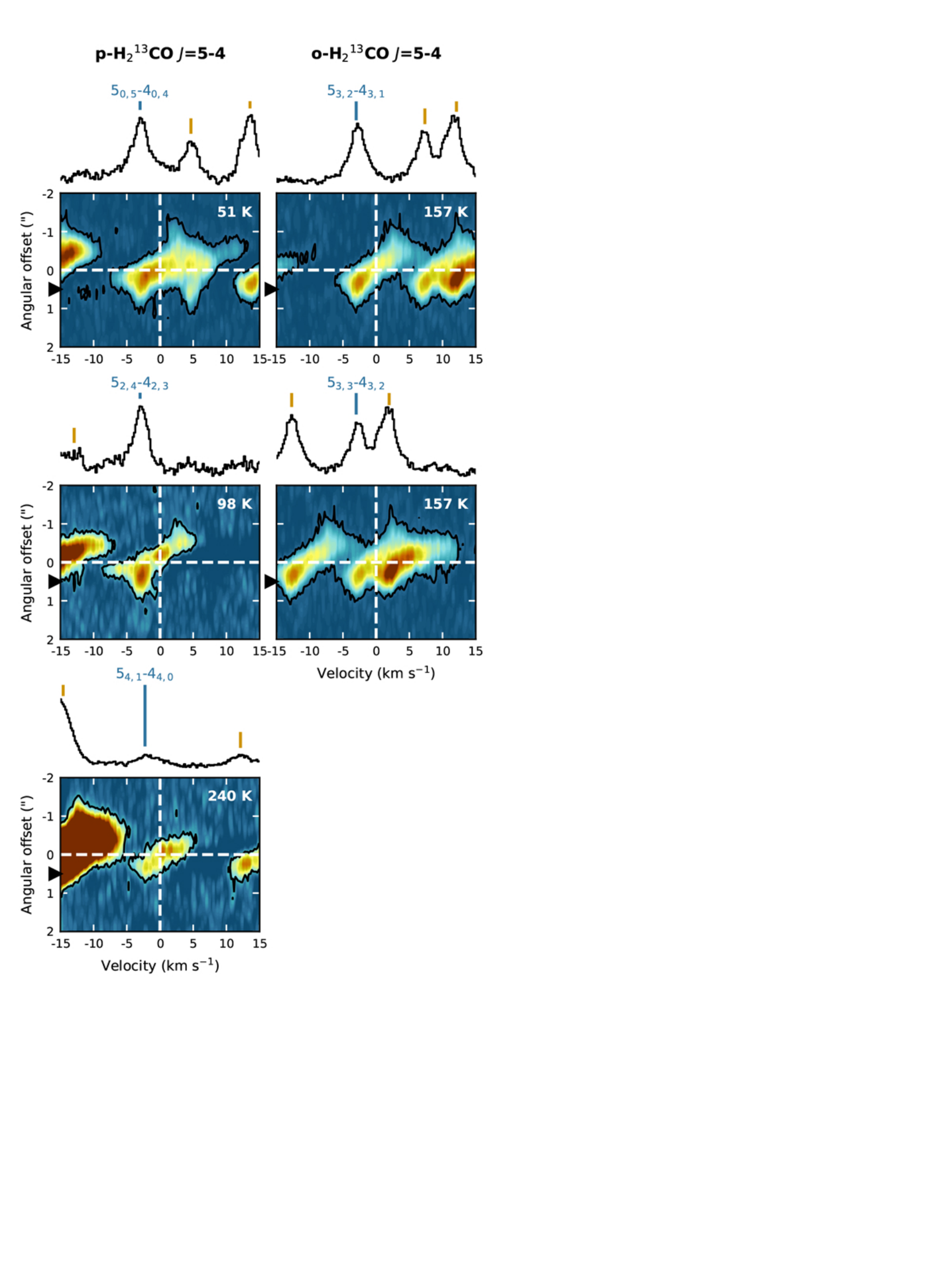}
\caption{As Fig.~\ref{fig:H2CO_pv}, but for H$_2^{13}$CO. }
\label{fig:H213CO_pv}
\end{figure}

\begin{figure}[!h]
\centering
\includegraphics[trim={0cm 0.4cm 0cm 0.8cm},clip]{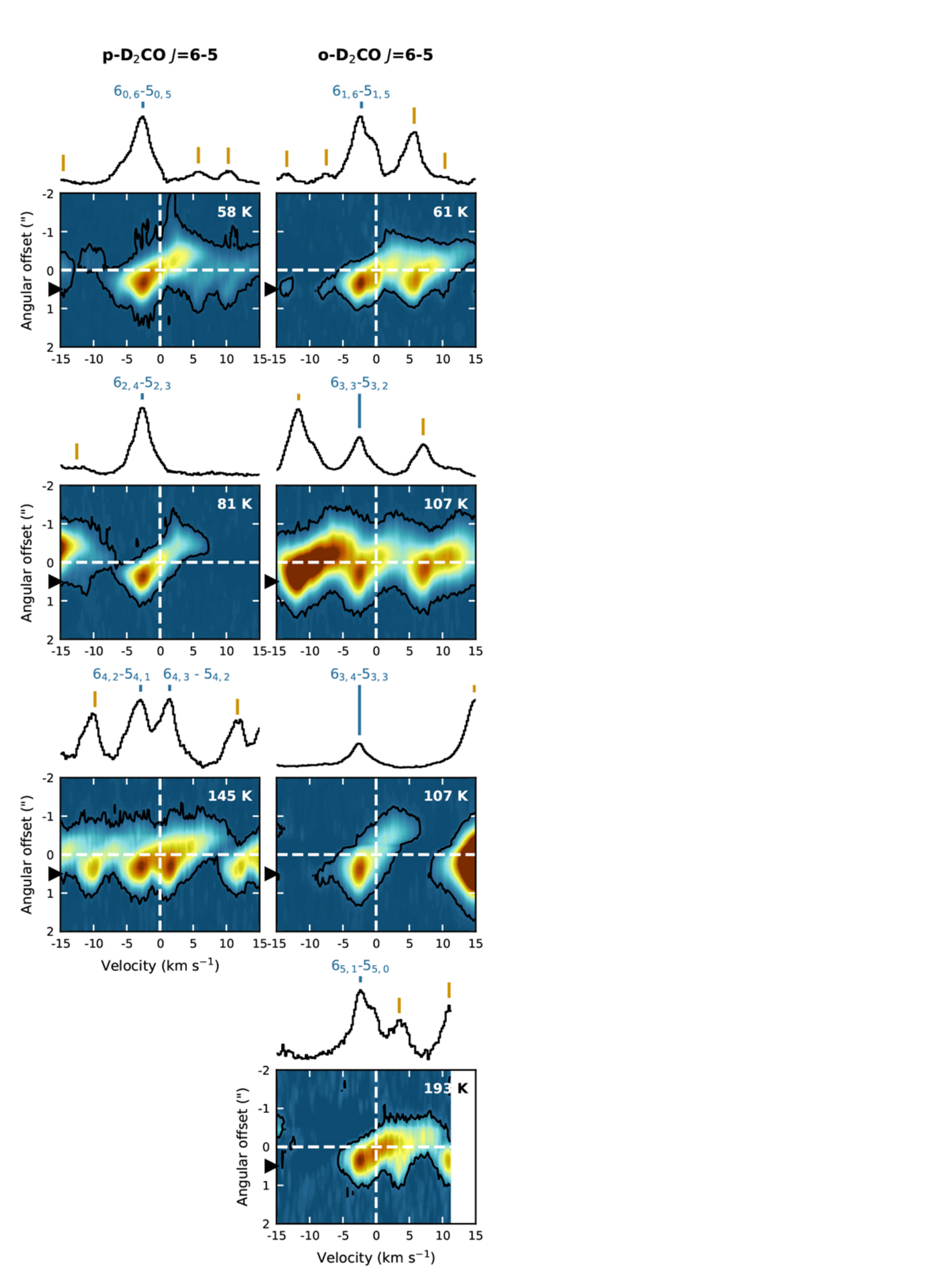}
\caption{As Fig.~\ref{fig:H2CO_pv}, but for D$_2$CO. }
\label{fig:D2CO_pv}
\end{figure}

Position-velocity diagrams for H$_2$CO, H$_2^{13}$CO and D$_2$CO are presented in Figs.~\ref{fig:H2CO_pv}, \ref{fig:H213CO_pv} and \ref{fig:D2CO_pv}, respectively. 

\section{Peak fluxes} \label{ap:Peakflux}

\begin{figure*}
\centering
\includegraphics[width=\textwidth]{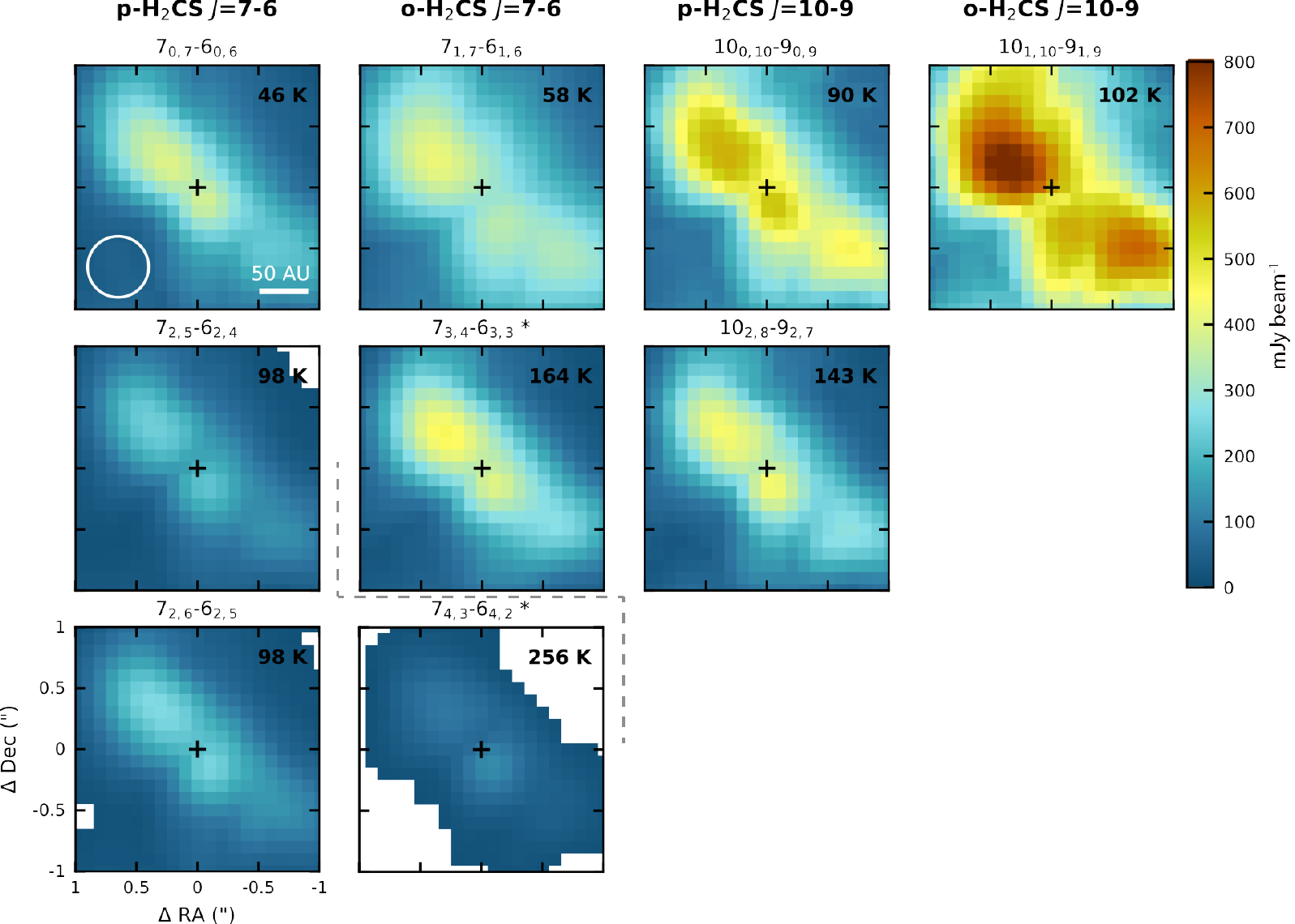}
\caption{Maps of the peak fluxes (moment 8) of the H$_2$CS lines. Only lines for which reliable peak fluxes could be extracted in the center, i.e., that are not too blended, are shown. The transitions are denoted above the panels and the corresponding upper level energies are listed in the top right corner of each panel. Asterisks ($\ast$) behind the transition indicate that the flux is coming from two lines with the same $J$ and $K_{\rm{a}}$-level located at the same frequency (see Table~\ref{tab:Lineoverview} for details). All panels are shown at the same intensity (color) scale for comparison. Pixels with no emission $>$3$\sigma$ are masked out. The continuum peak position is marked with a black cross and the beam is depicted in the lower left corner of the \textit{top left panel}.}
\label{fig:Peakflux_H2CS}
\end{figure*}

\begin{figure*}
\centering
\includegraphics[width=\textwidth]{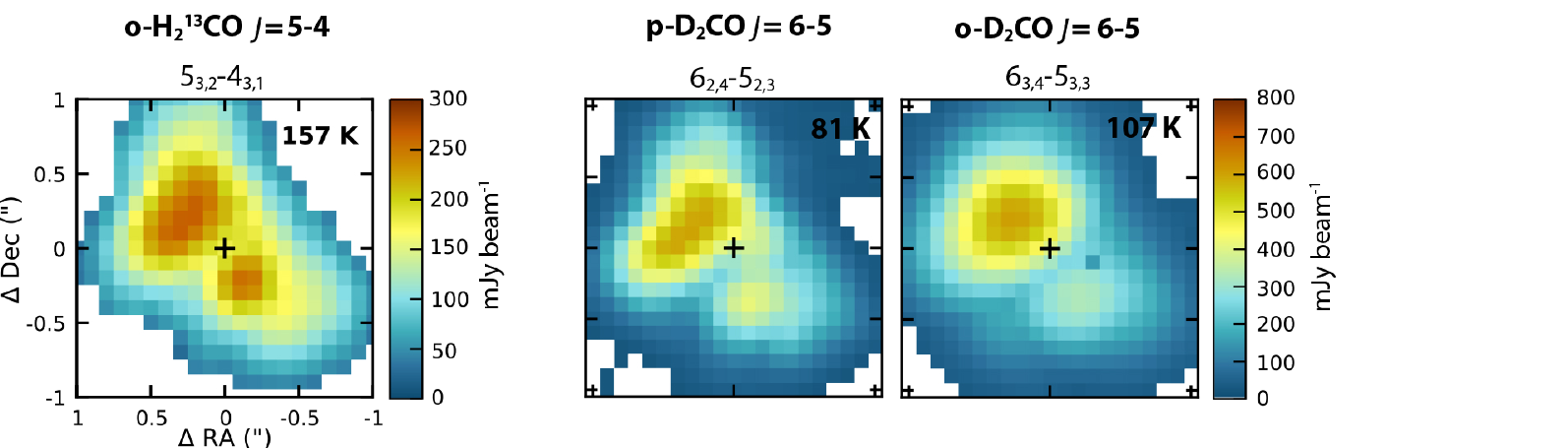}
\caption{As Fig.~\ref{fig:Peakflux_H2CS}, but for H$_2^{13}$CO (\textit{left panel}) and D$_2$CO (\textit{middle and right panel}).  }
\label{fig:Peakflux_H2CO}
\end{figure*}

Maps of the peak fluxes (moment 8 maps) are shown in Fig.~\ref{fig:Peakflux_H2CS} for H$_2$CS and in Fig.~\ref{fig:Peakflux_H2CO} for H$_2^{13}$CO and D$_2$CO. Lines that are too blended in the central region to extract reliable peak fluxes are excluded.  

\section{RADEX results for H$_2$CS} 

\begin{figure*}
\centering
\includegraphics[width=\textwidth,trim={0cm 5.5cm 0cm .7cm},clip]{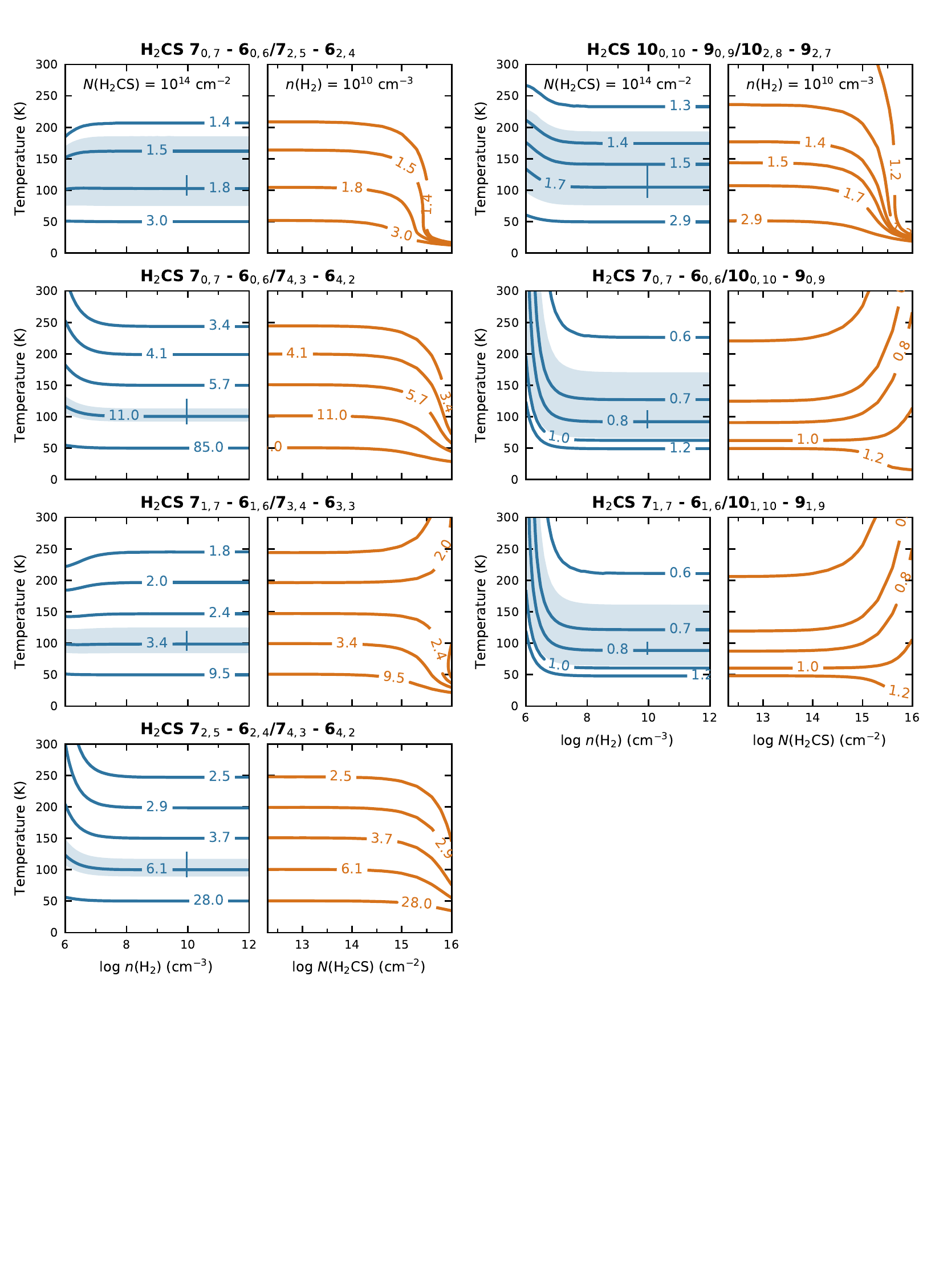}
\caption{Temperature sensitivity of the H$_2$CS line ratios as a function of H$_2$ density (blue lines) and H$_2$CS column density (orange lines). For the density plots, a H$_2$CS column density of $10^{14}$ cm$^{-2}$ is adopted for both o-H$_2$CS and p-H$_2$CS, and a H$_2$ density of $10^{10}$ cm$^{-3}$ is used for the column density plots. The line ratio is listed above the panels and contours are chosen such to mark the ratios at temperatures of 50, 100, 150, 200 and 250~K. The blue shaded area shows the change in derived temperature if the 100 K ratio changes by 20\%. The blue vertical line represents the typical 3$\sigma$ error on the temperature based on the observed uncertainty in the line ratio solely due to the rms in the images. The 7$_{0,7}$ -- 6$_{0,6}$/7$_{2,5}$ -- 6$_{2,4}$ ratio is equal to the 7$_{0,7}$ -- 6$_{0,6}$/7$_{2,6}$ -- 6$_{2,5}$ ratio (not shown) and the 7$_{2,5}$ -- 6$_{2,4}$/7$_{4,3}$ -- 6$_{4,2}$ ratio is equal to the 7$_{2,6}$ -- 6$_{2,5}$/7$_{4,3}$ -- 6$_{4,2}$ ratio (not shown). } 
\label{fig:RADEX}
\end{figure*}

\begin{figure}
\centering
\includegraphics[width=0.95\textwidth,trim={0cm 16.5cm 0cm 1cm},clip]{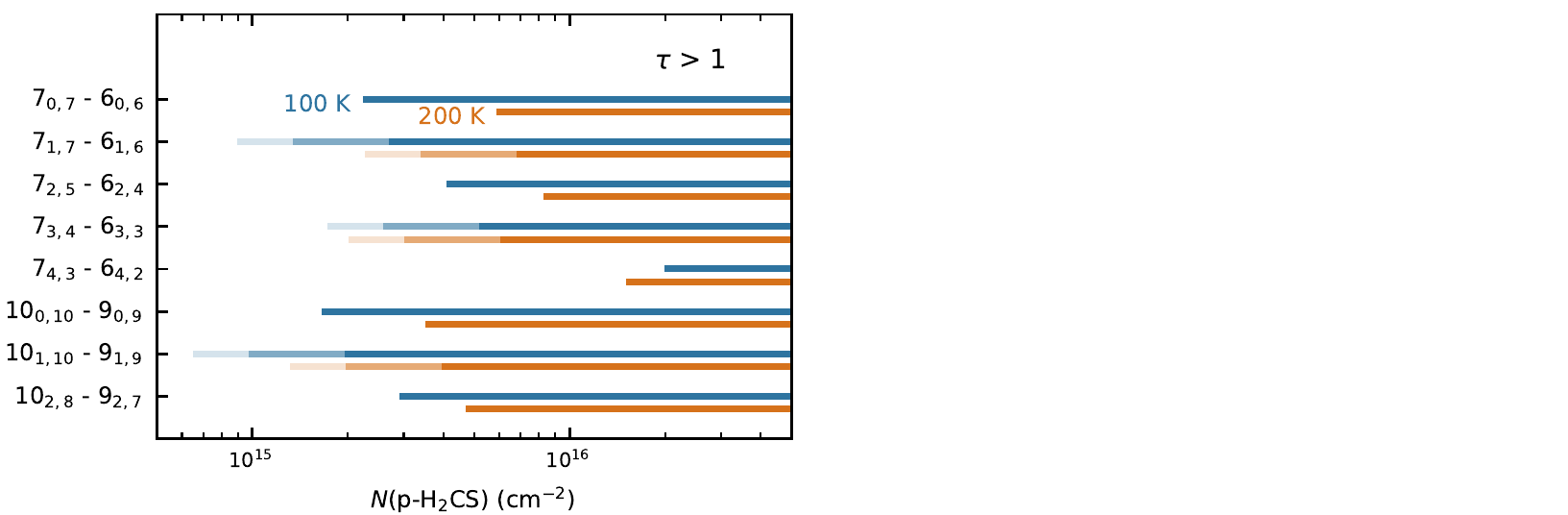}
\caption{Column densities (expressed in p-H$_2$CS column density) at which the H$_2$CS lines are optically thick at 100 K (blue) and 200 K (orange) based on the non-LTE RADEX calculation. For the ortho transitions, solid dark lines assume an ortho-to-para ratio of 1, and the shaded lines assume ortho-to-para ratios of 2 and 3 (lightest shade). }
\label{fig:tau_H2CS}
\end{figure}

H$_2$CS line ratios as function of temperature and density or temperature and column density derived from a grid of non-LTE radiative transfer calculations with RADEX are presented in Fig.~\ref{fig:RADEX}. Figure~\ref{fig:tau_H2CS} shows the column densities at which the different H$_2$CS lines become optically thick based on the same radiative transfer models.  

\section{Radial temperature profiles} \label{ap:Temperature}

\begin{figure}
\centering
\includegraphics[width=.94\textwidth,trim={0cm 10.6cm 0cm .8cm},clip]{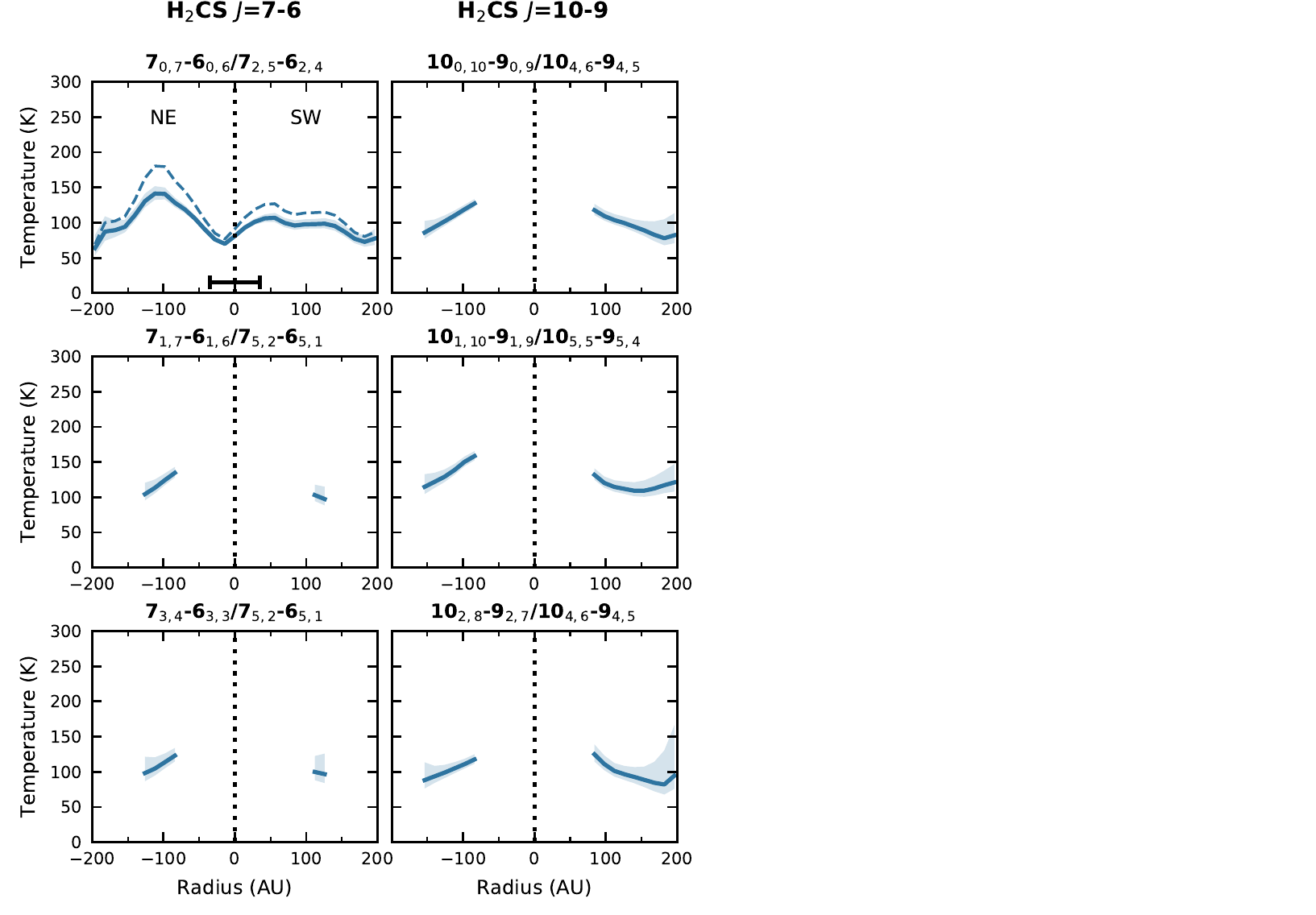}
\caption{Radial temperature profiles derived from H$_2$CS line ratios (listed above the panels) for which no collisional rate coefficients are available. The temperature profiles are taken along the major axis of the disk-like structure (PA = 45$^\circ$) and calculated assuming LTE. As a reference, the non-LTE result is shown as well for the 7$_{0,7}$ -- 6$_{0,6}$/7$_{2,5}$ -- 6$_{2,4}$ transition (dashed line in \textit{top left panel}). The shaded area represents the 1$\sigma$ uncertainty based on statistical errors. The horizontal bar in the \textit{top left panel} marks the beam size.}
\label{fig:Tradial-major_H2CS_LTE}
\end{figure}

\begin{figure*}
\centering
\includegraphics[width=\textwidth,trim={0cm 14cm 0cm .9cm},clip]{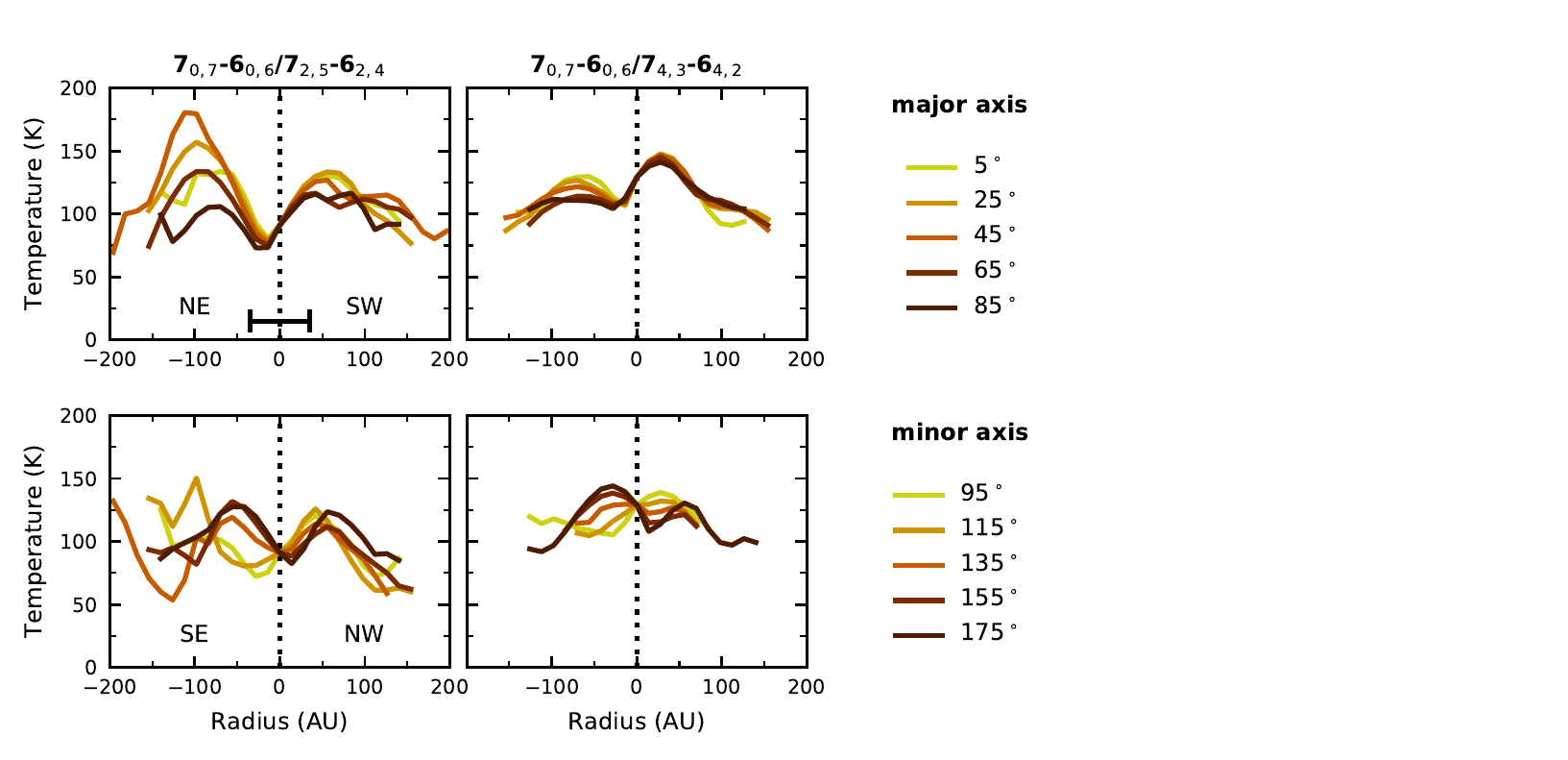}
\caption{Radial temperature profiles derived from two H$_2$CS line ratios (listed above the panels) along different position angles (different colors) using the non-LTE RADEX results. A PA of 45$^\circ$ corresponds to the major axis of the disk-like structure as presented in Fig~\ref{fig:Tradial-major_H2CS}. The horizontal bar in the \textit{top left panel} marks the beam size. }
\label{fig:Tradial_allmajor_H2CS}
\end{figure*}

Figure~\ref{fig:Tradial-major_H2CS_LTE} shows radial temperature profiles derived from H$_2$CS line ratios including transitions for which no collisional rate coefficients are available. These temperatures therefore result from an LTE calculation. Figure~\ref{fig:Tradial_allmajor_H2CS} presents the radial temperature profiles based on the 7$_{0,7}$ -- 6$_{0,6}$/7$_{2,5}$ -- 6$_{2,4}$ and 7$_{0,7}$ -- 6$_{0,6}$/7$_{4,3}$ -- 6$_{4,2}$ ratios (Fig.~\ref{fig:Tradial-major_H2CS} panels \textit{a} and \textit{c}) for different position angles. 

\end{appendix}

\end{document}